\documentclass[pra,twocolumn,showpacs,preprintnumbers,superscriptaddress]{revtex4-2}
\usepackage{xcolor}
\usepackage{times}
\usepackage{bm}
\usepackage{float}
\usepackage{graphicx}
\usepackage{amsbsy}
\usepackage{amsmath}
\usepackage{amsfonts}
\usepackage{amsthm}
\usepackage{placeins}

\begin{document}
	
	\allowdisplaybreaks
	
	\theoremstyle{plain}
	\newtheorem{theorem}{Theorem}
	\newtheorem{lemma}[theorem]{Lemma}
	\newtheorem{corollary}[theorem]{Corollary}
	\newtheorem{proposition}[theorem]{Proposition}
	\newtheorem{conjecture}[theorem]{Conjecture}
	
	\theoremstyle{definition}
	\newtheorem{definition}[theorem]{Definition}
	
	\theoremstyle{remark}
	\newtheorem*{remark}{Remark}
	\newtheorem{example}{Example}
	\title{Strength of the non-locality of two-qubit entangled states and its applications}
	\author{Anuma Garg, Satyabrata Adhikari}
	\email{anumagarg\_phd2k18@dtu.ac.in, satyabrata@dtu.ac.in} \affiliation{Delhi Technological University, Delhi-110042, Delhi, India}
	
	\begin{abstract}
		%\centerline{Abstract}
		Non-locality is a feature of quantum mechanics that cannot be explained by local realistic theory. It can be
		detected by the violation of Bell's inequality. In this work, we have considered the evaluation of Bell's inequality with the help of the XOR game. In the XOR game, a two-qubit entangled state is shared between the two distant players. It may generate a non-local correlation between the players which contributes to the maximum probability of winning of the game. We have aimed to determine the strength of the non-locality through XOR game. Thus, we have defined a quantity $S_{NL}$ called the strength of non-locality, purely on the basis of the maximum probability of winning of the XOR game. We have also derived the relation between the introduced quantity $S_{NL}$ and the quantity $M$ introduced in \cite{horo3}, to study the non-locality of a two-qubit entangled state problem in depth. The quantity $M$ may be defined as the sum of the two largest eigenvalues of the correlation matrix of the given entangled state and it determines whether the given entangled state under probe is non-local. Further, we have explored the non-locality of any two-qubit entangled state, whose non-locality cannot be detected by the CHSH inequality. 
		%but the corresponding CHSH witness operator does or does not detect the entangled state $\rho_{AB}^{ent}$.
		Interestingly, we have found that the newly defined quantity $S_{NL}$  fails to detect non-locality for the entangled state, when the witness operator corresponding to $CHSH$ operator cannot detect the entangled state. To overcome this problem, we have modified the definition of the strength of non-locality and have shown that the modified definition may detect the non-locality of such entangled states, which were earlier undetected by $S_{NL}$. Furthermore, we have provided two applications of the strength $S_{NL}$ of the non-locality such as (i) establishment of a link between the two-qubit non-locality determined by $S_{NL}$ and the three-qubit non-locality determined by the Svetlichny operator and (ii) determination of the upper bound of the power of the controller in terms of $S_{NL}$ in controlled quantum teleportation.
	\end{abstract}
	\maketitle

	\section{introduction}
	In 1964, J. S. Bell \cite{bell} derived a criterion to detect the non-local correlation that may exist in Einstein, Podolski, and Rosen (EPR) pair of particles. His work proved that the predictions of quantum mechanics are incompatible with the local realistic theory. Bell's criterion for detecing non-locality can be expressed in terms of a mathematical inequality, which is popularly known as Bell's inequality \cite{chsh,nielsen,wilde}, derived using the local-realism principle. Thus, any classical system making local choices will produce a classical correlation satisfying this inequality. In the late 1960s, many experiments were performed to show the violation of Bell's inequality for the EPR pair, but none were successful. Experiment performed by Alain Aspect et al.'s successfully shows the violation of Bell's Inequality \cite{aspect1,aspect2}. After Bell's seminal work, many studies were devoted to non-locality. The study of non-locality is relevant for many reasons. One is that it can be used as a resource for the development of device-independent quantum information processing \cite{vicente}. A few other reasons that may attract the study of non-locality is that it may have much application in a variety of quantum information processing tasks such as self-testing \cite{yang,bamps}, secure communication \cite{ekert}, randomness certification \cite{pironio}, and distributed computing \cite{abelson}. In a recent work, a marginal problem has been studied in the context of the computation of Bell inequalities \cite{glable}.\\ %Lately, non-locality is studied and implemented in real physical systems also \cite{moha1,moha2,moha3}. \\
	Detection of an observed non-local correlation is one of the prime problems in the study of non-locality. The foremost tool to detect non-locality is Bell's inequality \cite{bell}, which we have discussed in the previous paragraph, and it may be considered the standard approach for detecting non-locality. 
	The violation of Bell's inequality may indicate the presence of a non-local feature in a two-qubit state described by the density operator $\rho_{AB}$. Therefore, if any two-qubit state $\rho_{AB}$ violates Bell's inequality, then the state may exhibit non-local correlation, and thus, the state can be identified as an entangled state. But the converse of the statement is not valid. This means that there exists a two-qubit entangled state that may satisfy Bell's inequality. This shows that although, there is a connection between quantum entanglement and non-locality \cite{rahman,rahman 2}, conceptually they are very much distinct \cite{werner}. These two counterintuitive features of quantum mechanics can be used as a physical resource to enhance our computational power \cite{brito}. Thus, detecting these quantum mechanical features before using them as a resource is necessary. Along this line of research, I. S. Eliens et al. \cite{eliens} have studied the non-locality detection problem and represent it as a tensor network problem. In \cite{xguo}, the generalized $R$-matrix has been used to study the non-locality and entanglement of the three-qubit state. In \cite{rahman}, uncertainity induced non-locality measure has been used to detect the non-locality of two-qubit state. Further, the classification and quantification of a pure three-qubit state have been studied using the concurrence of a generic two-qubit pure state \cite{xguo1}. The witness operator method may also be used to detect non-locality, and it is very useful because it can be implemented in the experiment.\\ 
	In the last few years, testing of Bell's inequality has been viewed as a Bell game \cite{scarani}. In this game, Alice and Bob may be considered as players, and Charlie acts as a referee or verifier. There are many rounds of the game, and in each round, Charlie, who acts as a verifier, sends a query (input) to other members, Alice and Bob. They will have to send an answer (output) to Charlie. Before starting the game, the following assumptions are made: (i) Players know the set of possible queries, (ii) Players know the rules of the game (iii) Players know the common strategy in deciding the process in each round of the game. Here we can consider an entangled state as a resource that may be used in these processes. Therefore, in the perspective of a game, Bell locality may be defined as the process by which the output generated by each player is independent of the input of other players. Thus, if there is any correlation found between the players, then it is due to the presence of correlation in the shared entangled state. When this definition of Bell's locality does not hold, then we can talk of Bell's non-locality. Initially, Bell constructed the inequality in which two parties are there in the composite system and each party measures dichotomic observables in two different settings. Later, researchers have started generalising the Bell's non-locality with $N$ parties, $k$ measurement settings, and $d$ outcomes of the measurement \cite{mermin,belinsi,zukowski,kaszlikowski,collins,werner12}.
	It has been observed that two or more different non-local quantum behaviors may be responsible for the maximal violation of a Bell's inequality. However, the extremal quantum behavior can be realized by a unique (up to unitary equivalence) quantum representation \cite{franz}. The non-local correlation that  violate the Bell's inequalities maximally by unique quantum behaviors have been studied in \cite{acin12}. These Bell's inequalities are maximally violated by non-maximally entangled states also, thus showing that these state are necessary to characterize the boundary of the quantum region. The non-local correlation characterised by Bell's inequalities could be used as a resource for quantum optics, quantum computation, and quantum information. In this direction, Obada et al. \cite{moha1} have studied the link between non-locality and entanglement and have shown that the entangled state may possess the phenomenon of sudden death of entanglement and non-locality under the effect of thermal noise. The influence of the dissipation rate of the dissipative system on the quantum correlation has been studied in \cite{moha2}, using the Hilbert–Schmidt distance and Bell’s inequality correlations. They found that the quantum correlation can be enhanced for some specific values of the dipole–dipole interaction. In another work \cite{moha3}, it has been shown that the Bell's non-locality can be enhanced when the two-mode parametric amplifier cavity is initially prepared in the coherent states.\\ 	
	It is known that in any theory, the degree of steering is an equally important part as the uncertainty principle to measure the degree of non-locality \cite{rama1}. But J. Oppenheim and S. Wehner \cite{jon} have used the uncertainty principle alone to establish the relation between the maximum probability of winning of the XOR game and the expectation value of the Bell-CHSH operator with respect to the shared state between the players. Thus they have shown that the degree of non-locality can be determined by the uncertainty principle alone. Therefore, one may ask whether only one factor i.e., uncertainty principle is enough to measure the degree of non-locality for all non-local games. The answer is negative because R. Ramanathan et al. \cite{rama1} have shown that non-local games exist where the uncertainty principle and the degree of steering are needed to measure the degree of non-locality. But in particular, the degree of non-locality for the XOR game can be measured using the uncertainty principle alone. In the literature, there is a related work \cite{zhen} where it has been shown that some points that cannot maximize any XOR game lie on the quantum boundary.\\
	The main motivation of this work is to investigate the following question: If the Bell's inequality, the quantity $M$, and the maximum probability of winning of XOR game fail to determine the non-locality of an entangled state, and if we further restrict the usage of the filtering operation, then can we measure the strength of the non-locality by any other means?\\  
	To address the above stated question, we first consider the evaluation of Bell's inequality as an XOR game. The relation established in \cite{jon} suggests that if Bell's inequality is violated, then the maximum probability of winning is greater than $\frac{3}{4}$. Thus, there is a relation between the non-locality of the shared state and the maximum probability of winning of the game. We found that there exists an entangled shared state with which if players played the game, then the probability of winning the game may be less than or equal to $\frac{3}{4}$. This indicates that the XOR game may be won by adopting any local realistic theory, but this is not the case. We have investigated this loophole and tried to fix it by defining the strength of the non-locality through the maximum probability of winning of the game.\\
	This work can be distributed in different sections as follows: In section II, we have revisited the non-locality of the two-qubit entangled state $\rho^{ent}_{AB}$ by defining the strength of non-locality. We have established the relation between the maximum probability of winning and the CHSH witness operator and then studied the strength of non-locality by relating it with the expectation value of the CHSH witness operator. Moreover, in this section, we have studied the relation between $S_{NL}(\rho^{ent}_{AB})$ and $M(\rho^{ent}_{AB})$, which are considered as the two measures of non-locality of two-qubit entangled state $\rho^{ent}_{AB}$. In section III, we consider the optimal witness operator to study the strength of the non-locality. In section IV, we have provided the explicit expression of the strength of the non-locality in terms of measurement parameters and state parameters. Section V discusses the application of $S_{NL}(\rho^{ent}_{AB})$ to determine whether the particular classes of the three-qubit pure state are genuine non-local or not. Also, we have shown how $S_{NL}(\rho^{ent}_{AB})$ will be used to find out the upper bound of the power of the controller in controlled teleportation. Lastly, we end up with a conclusion. 
	\section{Revisiting the non-locality of two-qubit system}
	\noindent In a two-player Bell test game \cite{brunner}, the players may be referred to as Alice and Bob who are far apart from each other. Each player will receive a query (input) and will have to provide an answer (output). The game may be repeated many rounds. The players are allowed to prepare a common strategy before the game but after the game starts, the players are not allowed to communicate with each other. The rules of the game and the list of possible queries are known in advance. If the rule is set in a way that the players must produce different answers if both receive a query “1” and otherwise, the answer is same, then the game cannot be trivially won with a list of pre-determined answers. With respect to the defined game, Bell locality means that the process by which both the players generate the output without considering the other player’s input. That is, if any correlations generated between the players then this is due to a shared resource. The Bell non-locality came into the picture when Bell locality doesn't hold. Bell non-locality can be demonstrated by the violation of Bell-CHSH inequality. Generally, it has been shown by R. Horodecki et al. \cite{horo3} that any two-qubit state described by the density matrix $\rho_{AB}$ violates CHSH inequality if and only if $M(\rho_{AB})>1$. Here, $M(\rho_{AB})$ is defined as 
	\begin{eqnarray}
		M(\rho_{AB})=u_{1}+u_{2}
		\label{mrho}
	\end{eqnarray}
	where $u_{1}$ and $u_{2}$ are the two maximum eigenvalues of $T^{\dagger}T$. T denotes the $ 3 \otimes 3$ correlation matrix of $\rho_{AB}$, and its entries $t_{ij}$ can be calculated by 
	\begin{eqnarray}
		t_{ij}={Tr\rm}[\rho_{AB}(\sigma_{i}\otimes\sigma_{j})],~~ i,j=\{1,2,3\}
	\end{eqnarray}
	In this section, we revisit the non-locality of a two-qubit entangled state $\rho^{ent}_{AB}$ by introducing a measure of the strength of the non-locality of $\rho^{ent}_{AB}$. The motivation of this section is to develop a measure that may detect the non-local nature of the given entangled state $\rho^{ent}_{AB}$, which is neither detected by Bell-CHSH inequality (for a particular setting) nor detected by any general setting described by the criterion $M(\rho^{ent}_{AB})>1$. 
	\subsection{A definition of the strength of the non-locality of two-qubit entangled state}
	In this subsection, we will define the strength of the non-locality of two-qubit entangled state $\rho_{AB}^{ent}$ in terms of the maximum probability of winning of a game played between two distant players which are sharing an entangled state $\rho_{AB}^{ent}$.\\ 
	Let us consider an XOR game played between two distant players Alice (A) and Bob (B) \cite{jon,archan}. In this game, the winner is decided by the XOR of the answers $a \oplus b = a+b~(mod 2)$, where $a,b\in \{0,1\}$ and it denotes the answers given by the players A and B, when the referee asks them randomly selected questions $(s,t)\in S \times T$, where $S$ and $T$ denote finite non-empty sets. The winning condition of the game may be expressed in terms of the predicate given by
	\begin{eqnarray}
		V(a \oplus b/s,t)=1,~~\text{if and only if}~ a \oplus b =s.t  
	\end{eqnarray}
	The players A and B obtain outcomes (answers) $a$ and $b$ after performing measurement operators $A_{s}^{a}$ and $B_{t}^{b}$ on their respective qubits. Here, we may consider $s$ and $t$ as the corresponding measurement settings. The measurement operators $A_{s}^{a}$ and $B_{t}^{b}$ may be expressed in terms of the observables as
	\begin{eqnarray}
		&&A_{s}^{a}=\frac{1}{2}(I+(-1)^{a}A_{s})\nonumber\\&&
		B_{t}^{b}=\frac{1}{2}(I+(-1)^{b}B_{t})
	\end{eqnarray}
	The operators $A_{s}$ and $B_{t}$ are given by
	\begin{eqnarray}
		&&A_{s}=\sum_{j}a_{s}^{(j)}\Gamma_{j}\nonumber\\&&
		B_{t}=\sum_{j}a_{s}^{(j)}\Gamma_{j}
	\end{eqnarray}
	where $\vec{a}_{s}=(a_{s}^{(1)},a_{s}^{(2)},....,a_{s}^{(N)})\in R^{N}$ and $\vec{b}_{t}=(b_{t}^{(1)},b_{t}^{(2)},....,b_{t}^{(N)})\in R^{N}$ denote real unit vectors of dimension $N=min\{|S|,|T|\}$. $\Gamma_{1},\Gamma_{2},......,\Gamma_{N}$ are the anti-commuting generators of a Clifford algebra.\\
	In particular, we may consider four-dimensional space spanned by Pauli basis $\{I, \sigma_{x}, \sigma_{y}, \sigma_{z}\}$. In four-dimensional space, the two distant players may share the state $\rho_{AB}$, which is given by \cite{luo} 
	\begin{equation}
		\rho_{AB}= \frac{1}{4}[I\otimes I+ \overrightarrow{a}.\overrightarrow{\sigma}\otimes I +I \otimes \overrightarrow{b}.\overrightarrow{\sigma}+ \sum c_{j} \sigma_{j} \otimes \sigma_{j}]
		\label{gen2qbitst22}
	\end{equation} 
	where $\vec{a}=(a_{1},a_{2},a_{3})\in R^{3}$, $\vec{b}=(b_{1},b_{2},b_{3})\in R^{3}$, $c_{i}\in R$ and $\sigma_{i}$ denote the Pauli matrices.\\
	If we assume that the two distant players, A and B, play the game using the shared state $\rho_{AB}$, then the maximum probability $P^{max}$ of winning the game overall strategy is given by \cite{jon} 
	\begin{eqnarray}
		P^{max}&=&\frac{1}{2}[1+\frac{\langle B_{CHSH}\rangle _{\rho_{AB}}}{4}]
		\label{pmax}
	\end{eqnarray}
	where $\langle B_{CHSH}\rangle _{\rho_{AB}}={Tr\rm}[(A_{0}\otimes B_{0}+A_{0}\otimes B_{1}+A_{1}\otimes B_{0}-A_{1}\otimes B_{1})\rho_{AB}]$ denotes the expectation value of the Bell operator $B_{CHSH}$ with respect to the state $\rho_{AB}$. Since, the maximum probability of winning the game depends on the expectation value of the Bell operator $B_{CHSH}$, so $P^{max}$ is somehow related to the non-locality of the state $\rho_{AB}$. Thus, to determine the non-locality of any arbitrary two-qubit state $\rho_{AB}$, we define here the strength of the non-locality. The strength of the non-locality of $\rho_{AB}$ denoted by $S_{NL}(\rho_{AB})$ in terms of $P^{max}$ may be defined as 
	\begin{equation}
		S_{NL}(\rho_{AB})=max \{P^{max}-\frac{3}{4},0\}
		\label{snl22}
	\end{equation}
	Therefore, $S_{NL}(\rho_{AB})$ can be considered as the quantifier of the strength of the non-locality for any theory, and it can be calculated by calculating $P^{max}$  for different theories such as (i) classical theory, (ii) theory based on quantum mechanics, and (iii) for any non-signaling theory. For any classical theory, $P^{max}\leq \frac{3}{4}$ and hence $S_{NL}(\rho_{AB})=0$. For quantum mechanical theory and for non-signaling correlation, we have $P^{max} > \frac{3}{4}$   and thus $S_{NL}(\rho_{AB})\neq 0$.\\
	Furthermore, we can consider the situation where the players performed their measurements in different measurement settings, such as measurements performed along $xy-$, $xz-$, and $yz-$ planes. In this scenario, the maximum probability of winning the game depends upon the expectation value of the Bell operators in different planes. To further illuminate this point, consider the Bell operators $B_{xy}$, $B_{xz}$, and $B_{yz}$ in $xy-$, $xz-$, and $yz-$ planes. In these planes, the maximum probability of winning of winning the game is denoted by $P_{xy}$, $P_{xz}$, and $P_{yz}$, respectively. Therefore, the relation between the expectation value of the Bell operators defined in different planes with respect to the two-qubit quantum state described by the density operator $\rho_{AB}$ and the corresponding maximum probability of winning may be expressed as
	\begin{equation}
		P_{ij}^{max}=\frac{1}{2}[1+\frac{\langle B_{ij}\rangle_{\rho_{AB}}}{4}] ,i,j=x,y,z~~ \&~~  i\neq j
		\label{R1}
	\end{equation}   
	The Bell operators $B_{xy}$, $B_{xz}$, and $B_{yz}$ can be written in terms of the observables $\sigma_{x}$, $\sigma_{y}$, and $\sigma_{z}$ as \cite{hyllus}
	\begin{eqnarray}
		B_{ij}&=&\sigma_{i}\otimes \frac{\sigma_{i}+\sigma_{j}}{\sqrt{2}}+ \sigma_{i}\otimes \frac{\sigma_{i}-\sigma_{j}}{\sqrt{2}}\nonumber\\&+&
		\sigma_{j}\otimes \frac{\sigma_{i}+\sigma_{j}}{\sqrt{2}}- \sigma_{j}\otimes \frac{\sigma_{i}-\sigma_{j}}{\sqrt{2}},\nonumber\\&& i,j=x,y,z~~ \&~~ i\neq j
		\label{R2}
	\end{eqnarray}
	For the case discussed above, the strength of the non-locality $S_{NL}^{(ij)}(\rho_{AB})$ may be defined as
	\begin{eqnarray}
		S_{NL}^{(ij)}(\rho_{AB})=max\{P,0\}
		\label{snl23}
	\end{eqnarray}
	where $P=\{P^{max}_{xy}-\frac{3}{4}, P^{max}_{xz}-\frac{3}{4}, P^{max}_{yz}-\frac{3}{4}\}$.\\
	From the definition $S_{NL}^{(ij)}(\rho_{AB})$ given in (\ref{snl23}), it is clear that if $\rho_{AB}$ is an entangled state, and further if it satisfies the Bell-CHSH inequality in every $xy$, $yz$, $xz$ setting, then all quantities $P_{ij}^{max}-\frac{3}{4}, (i\neq j;~i,j=x,y,z)$ will be negative. Hence, the value of $S_{NL}^{(ij)}(\rho_{AB})$ for $i\neq j;~i,j=x,y,z$ will be equal to zero. This gives an absurd result, because $\rho_{AB}$ represents an entangled state. Thus, we can apply the definition (\ref{snl23}) only when at least one $i\neq j$; $(i,j=x,y,z)$ for which the quantity $P_{ij}^{max}-\frac{3}{4}$ is/are positive.\\
	\subsection{Dependence of the strength of non-locality on witness operator} 
	Let us consider the same game played between two distant partners, Alice and Bob, using a shared state between them. If the shared state is any entangled state described by $\rho^{ent}_{AB}$ and the maximum probability $P^{max}$ of winning the game using the shared state $\rho^{ent}_{AB}$ satisfies the inequality $P^{max} > \frac{3}{4}$, then as per the definition (given in the previous section) of the strength of the non-locality of $\rho^{ent}_{AB}$ could be non-zero. Otherwise, if the players are playing the game with the classical state shared between them, then $P^{max} \in [0,\frac{3}{4}]$ and then the strength of the non-locality will be equal to zero. This event may occur even if the players choose their measurement settings in different planes. In this perspective, we can ask the following questions: (i) whether there exists any entangled state shared between two players for which the maximum probability of winning the game lies between $0$ and $\frac{3}{4}$? and (ii) are we able to determine the strength of the non-locality in this case?\\ 
	To investigate the above questions, we first express the maximum probability $P^{max}$ of winning the game in terms of the expectation value of the witness operator with respect to the general two-qubit state described by the density operator $\rho_{AB}^{gen}$. Also, we find that when $\rho_{AB}^{gen}$ represents an entangled state which is not detected by the witness operator, then the maximum probability of winning the game lies between $0$ and $\frac{3}{4}$. On the contrary, if there exists any witness operator that detects the entangled state, then $P^{max}\geq \frac{3}{4}$.\\
	Let us now establish the relationship between the maximum probability of winning the game played using a two-qubit state $\rho_{AB}^{gen}$, and the expectation value of the witness operator with respect to the state $\rho_{AB}^{gen}$. The relationship may be stated as:\\ 
	\textbf{Result-1:-} If $\rho_{AB}^{gen}$ denotes any arbitrary two-qubit bipartite state shared between the two distant players Alice and Bob and $P^{max}$ denotes the maximum probability of winning the game overall strategy taken by them, then $P^{max}$ is given by 
	\begin{equation}
		P^{max}=\frac{3}{4}-\frac{{Tr\rm}[W_{CHSH}\rho_{AB}^{gen}]}{8}
		\label{Res1}
	\end{equation}
	where, $W_{CHSH}(=2I-B_{CHSH})$ denotes the witness operator.\\ %and $B_{CHSH}$ denote the Bell-CHSH operator \cite{nielsen,wilde}.\\ %(\cite{nielsen},\cite{wilde}). \\
	\textbf{Proof}:-
	If any bipartite two-qubit state $\rho_{AB}^{gen}$ is shared between the players Alice and Bob, then the maximum probability of winning the game is given by \cite{jon} 
	\begin{eqnarray}
		P^{max}&=&\frac{1}{2}[1+\frac{{Tr\rm}[B_{CHSH}\rho_{AB}^{gen}]}{4}] \nonumber \\
		&=&\frac{3}{4}-\frac{{Tr\rm}[W_{CHSH}\rho_{AB}^{gen}]}{8} 
		\label{witness1}		
	\end{eqnarray}
	In the second line of the proof, we have used $B_{CHSH}=2I-W_{CHSH}$. Hence proved.\\
	In the same spirit, we can relate the maximum probability of winning the game with the expectation value of the witness operator in different $xy$, $yz$, and $xz$ settings, such as
	\begin{eqnarray}
		P_{xy}^{max}=\frac{3}{4}-\frac{{Tr\rm}[W_{CHSH}^{xy}\rho_{AB}^{gen}]}{8} 
		\label{witness21}		
	\end{eqnarray}
	\begin{eqnarray}
		P_{yz}^{max}=\frac{3}{4}-\frac{{Tr\rm}[W_{CHSH}^{yz}\rho_{AB}^{gen}]}{8} 
		\label{witness22}		
	\end{eqnarray}
	\begin{eqnarray}
		P_{xz}^{max}=\frac{3}{4}-\frac{{Tr\rm}[W_{CHSH}^{xz}\rho_{AB}^{gen}]}{8} 
		\label{witness22}		
	\end{eqnarray}
	\subsubsection{Strength of the non-locality when two-qubit entangled state detected by the witness operator $W_{CHSH}$} 
	In this subsection, we will discuss the case when the witness operator detects the entangled state $\rho_{AB}^{ent}$ and then we show that the strength of the non-locality denoted by $S_{NL}(\rho_{AB}^{ent})$ can be determined in this case.\\ 
	Result-1 provides the relationship between the expectation value of the witness operator $W_{CHSH}$ with respect to any arbitrary two-qubit state $\rho_{AB}^{gen}$, and the maximum winning probability $P^{max}$. Therefore, the strength of the non-locality $S_{NL}(\rho^{gen})$ defined in (\ref{snl22}) may be re-expressed in terms of witness operator $W_{CHSH}$ as
	\begin{eqnarray}
		S_{NL}(\rho_{AB}^{gen})=max \{-\frac{{Tr\rm}[W_{CHSH}\rho_{AB}^{gen}]}{8},0\}
		\label{snl1}		
	\end{eqnarray} 
	%In general, the strength of the non-locality $S_{NL}$ of the state $\rho_{AB}$ may be defined in terms of any witness operator $W_{gen}$ as
	%\begin{eqnarray}
	%S_{NL}=max \{-Tr[W_{gen}\rho_{AB}],0\}
	%\label{snlgen}		
	%\end{eqnarray} 
	Let us discuss three cases when $\rho_{AB}^{gen}$ represents (i) a separable state, (ii) an entangled state not detected by $W_{CHSH}$, and (iii) an entangled state detected by $W_{CHSH}$.\\
	\textbf{Case-I:} If any separable state is described by the density operator $\rho_{AB}^{sep}$, then ${Tr\rm}[W_{CHSH}\rho_{AB}^{sep}]\geq 0$ and hence $P^{max}\leq \frac{3}{4}$. In this case, $S_{NL}(\rho_{AB}^{sep})=0$.\\
	\textbf{Case-II:} If the state $\rho_{AB}^{entnd}$ denotes an entangled state and it is not detected by witness operator $W_{CHSH}$, then also we obtain ${Tr\rm}[W_{CHSH}\rho_{AB}^{entnd}]\geq 0$ and hence $P^{max}\leq \frac{3}{4}$. In this case, the amount of non-locality of the state $\rho_{AB}^{entnd}$ can be estimated to be zero. Although the state $\rho_{AB}^{entnd}$ is an entangled state and thus may possess non-local properties but its non-locality may not be revealed by the non-local quantifier $S_{NL}$. Further, we may note that the state $\rho_{AB}^{entnd}$ may not be detected by $W_{CHSH}$, but there may exist other witness operators that may detect it, and in that case, it may be possible to quantify its non-locality through $S_{NL}$. \\
	\textbf{Case-III:} If the state $\rho_{AB}^{entd}$ is an entangled state and witness operator $W_{CHSH}$ detects it, then ${Tr\rm}[W_{CHSH}\rho_{AB}^{entd}]< 0$ and hence $P^{max}> \frac{3}{4}$. In this case, the amount of non-locality of  $\rho_{AB}^{entd}$ can be calculated by the formula $S_{NL}=-\frac{{Tr\rm}[W_{CHSH}\rho_{AB}^{entd}]}{8}$.\\
	Let us now take an example for \textbf{Case-III}. In this example, it is possible to show that the two-qubit state under investigation is a quantum correlated state, and thus the strength of non-locality of it can be determined. To proceed with our discussion,
	let us consider the two-qubit quantum state described by the density operator $\rho_{AB}^{(1)}$
	\begin{eqnarray}
		\rho_{AB}^{(1)}&=& \frac{1}{4}[I\otimes I+ 0.001\sigma_{x}\otimes I + 0.8 \sigma_{1} \otimes \sigma_{1}+ 0.89 \sigma_{2} \otimes \sigma_{2}\nonumber\\&-&0.9 \sigma_{3} \otimes \sigma_{3}] 
		\label{par2qbitst21}
	\end{eqnarray} 
	The state $\rho_{AB}^{(1)}$ is an entangled state.\\
	In this case, we can construct the witness operator $W_{CHSH}^{(1)}$ as 
	\begin{eqnarray}
		W_{CHSH}^{(1)}&=&2I \otimes I-A_{0}^{(1)}\otimes B_{0}^{(1)}+A_{0}^{(1)}\otimes B_{1}^{(1)}\nonumber\\&-& A_{1}^{(1)}\otimes B_{0}^{(1)}-A_{1}^{(1)}\otimes B_{1}^{(1)}
	\end{eqnarray}
	where 
	\begin{eqnarray}
		A_{0}^{(1)}&=& \sigma_{x}\nonumber \\
		A_{1}^{(1)}&=& \sigma_{y}\nonumber \\ 
		B_{0}^{(1)}&=&0.8\sigma_{x}+0.4\sigma_{y}+0.447\sigma_{z}\nonumber \\
		B_{1}^{(1)}&=&-0.4\sigma_{x}+0.8\sigma_{y}+0.447\sigma_{z}
	\end{eqnarray}
	Therefore, the expectation value of $W_{CHSH}^{(1)}$ with respect to the state $\rho_{AB}^{(1)}$ is given by 
	\begin{eqnarray}
		{Tr\rm}[W_{CHSH}^{(1)}\rho_{AB}^{(1)}]&=&-0.028 <0 
	\end{eqnarray}
	Hence, in this example, we can see the state $\rho_{AB}^{(1)}$ is detected as an entangled state by the witness operator $W_{CHSH}^{(1)}$. Thus, the strength of the non-locality of the state $\rho_{AB}^{(1)}$ can be calculated using (\ref{snl1}), and which is
	\begin{eqnarray}
		S_{NL}(\rho_{AB}^{(1)})= 0.0035
	\end{eqnarray}
	\subsubsection{Strength of the non-locality when the witness operator $W_{CHSH}$ does not detect the two-qubit entangled state} 
	Till now, we don't have sufficient information to make a definite conclusion about the non-locality of an entangled state described by the density operator $\rho_{AB}^{entnd}$, which is not detected by the witness operator $W_{CHSH}$. Let us take an example to understand what we mean to say: Consider the state $\rho_{AB}^{(2)}$, which is given by
	\begin{equation}
		\rho_{AB}^{(2)}= \frac{1}{4}[I\otimes I+ 0.7 \sigma_{1} \otimes \sigma_{1}+ 0.2 \sigma_{2} \otimes \sigma_{2}-0.5 \sigma_{3} \otimes \sigma_{3}]
		\label{gen2qbitentst}
	\end{equation} 
	It can be shown that the state $\rho_{AB}^{(2)}$ is an entangled state.\\ 
	Let us now consider the witness operator $W^{(2)}_{CHSH}$, which is given by  
	\begin{eqnarray}
		W_{CHSH}^{(2)}&=&2I\otimes I-A_{0}^{(2)}\otimes B_{0}^{(2)}+A_{0}^{(2)}\otimes B_{1}^{(2)}\nonumber\\&&-A_{1}^{(2)}\otimes B_{0}^{(2)}-A_{1}^{(2)}\otimes B_{1}^{(2)} 
		\label{witchsh2}
	\end{eqnarray}
	where $A_{0}^{(2)}$, $A_{1}^{(2)}$, $B_{0}^{(2)}$, $B_{1}^{(2)}$ are given by
	\begin{eqnarray}
		A_{0}^{(2)}&=&0.7\sigma_{x}+0.5\sigma_{y}+0.5099\sigma_{z}\nonumber \\
		A_{1}^{(2)}&=&0.7\sigma_{x}+0.5\sigma_{y}+0.5099\sigma_{z}\nonumber \\ 
		B_{0}^{(2)}&=&0.4\sigma_{x}+0.4\sigma_{y}+0.8246\sigma_{z}\nonumber \\
		B_{1}^{(2)}&=&0.5\sigma_{x}+0.3\sigma_{y}+0.812404\sigma_{z}
	\end{eqnarray}
	The expectation value of $W^{(2)}_{CHSH}$ with respect to the state $\rho_{AB}^{(2)}$ can be calculated as
	\begin{eqnarray}
		{Tr\rm}[W_{CHSH}^{(2)}\rho_{AB}^{(2)}] &=& 1.9845 \geq 0  
	\end{eqnarray}
	Thus, this example shows that there may exist entangled states which are not detected by $W_{CHSH}^{(2)}$ operator given in (\ref{witchsh2}), and from Result-1, we have remarked $P^{max}\leq \frac{3}{4}$. Hence, we conclude that there exist entangled states for which $P^{max}\leq \frac{3}{4}$. Therefore, for those entangled states which are not detected by $W_{CHSH}$, we find $S_{NL}=0$, and thus $S_{NL}$ is unable to measure the true strength of non-locality of such entangled states. This problem may be sorted out if we construct another witness operator that may detect such entangled states which are not detected by $W_{CHSH}$. Since, there does not exist any general relationship between the maximum probability $P^{max}$ and the expectation value of any arbitrary witness operator, it is not possible to define the strength of the non-locality in terms of any arbitrary witness operator. Therefore, we need to redefine the strength of the non-locality using a different approach.\\
	It is known from (\ref{Res1}) that if $W_{CHSH}$ fails to detect the entangled state $\rho_{AB}^{ent}$, then the value of the expression  $P^{max}-\frac{3}{4}$ will be negative. Thus, our idea is to calculate the upper bound of the expression $P^{max}-\frac{3}{4}$ and if we find that the calculated upper bound is positive, then we may infer that there may be a possibility to get the non-zero value of $S_{NL}(\rho_{AB}^{ent})$.
	%Thus, the strength of the non-locality $S_{NL}(\rho_{AB}^{ent})$ comes out to be zero. This value of $S_{NL}(\rho_{AB}^{ent})$ seems to be incorrect since $\rho_{AB}^{ent}$ is an entangled state. Therefore, it is necessary to investigate this fact and in doing so, we will take some different route. 
	To do this, recall (\ref{Res1}) and re-express it as
	\begin{eqnarray}
		{Tr\rm}[W_{CHSH}\rho_{AB}^{ent}]&=&6-8P^{max} 
		\label{witprob}
	\end{eqnarray}	
	We should note that in this scenario, it is assumed that $W_{CHSH}$ does not detect the state $\rho_{AB}^{ent}$ and thus ${Tr\rm}[W_{CHSH}\rho_{AB}^{ent}]\geq 0$, hence $P^{max}\leq \frac{3}{4}$. To proceed further, we need a result, which can be stated as\\
	\textbf{Result-2 \cite{lasserre}:} For any two Hermitian matrices $X$ and $Y$,
	we have
	\begin{eqnarray}
		\lambda_{min}(X){Tr\rm}(Y) \leq {Tr\rm}(XY) \leq \lambda_{max}(X){Tr\rm}(Y)
		\label{las}
	\end{eqnarray}	
	Let us now re-start with the quantity ${Tr\rm}[W_{CHSH}\rho_{AB}^{ent}(\rho_{AB}^{ent})^{T_{B}}]$, where $T_{B}$ denotes the partial transposition with respect to the subsystem $B$ and using the result given in (\ref{las}), we may get the following inequality
	\begin{eqnarray}
		{Tr\rm}[W_{CHSH}\rho_{AB}^{ent}(\rho_{AB}^{ent})^{T_{B}}]&\geq& \lambda_{min}((\rho_{AB}^{ent})^{T_{B}})\times \nonumber\\&&{Tr\rm}[W_{CHSH}\rho_{AB}^{ent}]
		\label{ineq23}
	\end{eqnarray}
	Using (\ref{witprob}) and (\ref{ineq23}), we get
	\begin{eqnarray}
		{Tr\rm}[W_{CHSH}\rho_{AB}^{ent}(\rho_{AB}^{ent})^{T_{B}}]&\geq& \lambda_{min}((\rho_{AB}^{ent})^{T_{B}}) \times\nonumber\\&&(6-8P^{max})
		\label{ineq24}
	\end{eqnarray}	
	If $\rho_{AB}^{ent}$ is a bipartite two-qubit entangled state, then $\lambda_{min}((\rho_{AB}^{ent})^{T_{B}})<0$, and
	its entanglement may be quantified by negativity, which may be defined as
	\begin{eqnarray}
		N(\rho_{AB}^{ent})= -2 \lambda_{min}((\rho_{AB}^{ent})^{T_{B}})
		\label{neg}
	\end{eqnarray}	
	Therefore, for the entangled state $\rho_{AB}^{ent}$, the inequality (\ref{ineq24}) reduces to
	\begin{eqnarray}
		&&{Tr\rm}[W_{CHSH}\rho_{AB}^{ent}(\rho_{AB}^{ent})^{T_{B}}]\geq -\frac{1}{2}N(\rho_{AB}^{ent})(6-8P^{max}) \nonumber\\&&
		\implies P^{max}-\frac{3}{4}\leq\frac{{Tr\rm}[W_{CHSH}\rho_{AB}^{ent}(\rho_{AB}^{ent})^{T_{B}}]}{4N(\rho_{AB}^{ent})} 
		\label{ineq25}				
	\end{eqnarray}
	The inequality (\ref{ineq25}) motivates us to re-define the strength of the non-locality $S_{NL}(\rho_{AB}^{ent})$ of the entangled state $\rho_{AB}^{ent}$ undetected by $W_{CHSH}$. Therefore, if the state $\rho_{AB}^{ent}$ is not detected by $W_{CHSH}$ and  then $S^{New}_{NL}(\rho_{AB}^{ent})$ may be defined as 
	\begin{eqnarray}
		S^{New}_{NL}(\rho_{AB}^{ent})&=&q(P^{max}-\frac{3}{4})+(1-q)K
		\label{snldef}
	\end{eqnarray}
	where $K=\frac{{Tr\rm}[W_{CHSH}\rho_{AB}^{ent}(\rho_{AB}^{ent})^{T_{B}}]}{4N(\rho_{AB}^{ent})}$ and $q~(0\leq q<1)$ is chosen in such a way that $S^{New}_{NL}(\rho_{AB}^{ent})>0$.\\
	The upper bound of $q$ can be obtained by employing the condition $S^{New}_{NL}(\rho_{AB}^{ent})>0$. Therefore, the upper bound of $q$ is given by
	\begin{eqnarray}
		q<\frac{K}{\frac{3}{4}-P^{max}+K}
		\label{qub}
	\end{eqnarray}
	To illustrate our result, let us consider the state described by the density operator $\rho_{AB}$
	\begin{eqnarray}
		\rho_{AB}=\begin{pmatrix} 
			x & 0& 0& 0\\
			0 & \frac{1}{3} & x & 0\\
			0 & x & \frac{1}{3} & 0\\
			0 & 0& 0& \frac{1}{3}-x \\
		\end{pmatrix},~~ 0\leq x \leq \frac{1}{3}
	\end{eqnarray}
	It can be easily verified that $\rho_{AB}$ is an entangled state for $x \in (0.167,0.333)$. Also, we found that for the same range of $x$, we have ${Tr\rm}[W_{CHSH}^{xy}\rho_{AB}]=2-4\sqrt{2}x \geq 0$. Therefore, the state $\rho_{AB}$ is undetected by the witness operator $W_{CHSH}^{xy}$.\\
	To calculate the strength of the non-locality of $\rho_{AB}$, we follow the definition (\ref{snldef}) and accordingly determine the following quantities,
	\begin{eqnarray}
		&&P^{max}-\frac{3}{4}= 4\sqrt{2}x-2\nonumber\\&&
		K=\frac{1-2(1+\sqrt2)x+6x^2}{2(\sqrt(72x^2-12x+1)-1)}
		\label{pk}
	\end{eqnarray}
	Therefore, using (\ref{qub}), we find that
	\begin{eqnarray}
		q< [0.55,1], ~~\text{when}~ x\in (0.1667,0.333)
		\label{ex-1}
	\end{eqnarray}
	Therefore, the strength of the non-locality of the state $\rho_{AB}$ is given by 
	\begin{eqnarray}
		S_{NL}^{New}(\rho_{AB})=q(P^{max}-\frac{3}{4})+(1-q)K,~0<q<0.55 \nonumber \\
	\end{eqnarray}
	where the expressions $P^{max}-\frac{3}{4}$ and $K$ are given in (\ref{pk}). The value of $S_{NL}^{New}(\rho_{AB})$ for x and q satisfying (\ref{ex-1}) are shown in figure-1. So, by this process, we are able to calculate the strength of non-locality for the entangled states probabilistically which are not detected by $W_{CHSH}$.  
	\begin{figure}[h!]
		\centering
		\includegraphics[scale=0.55]{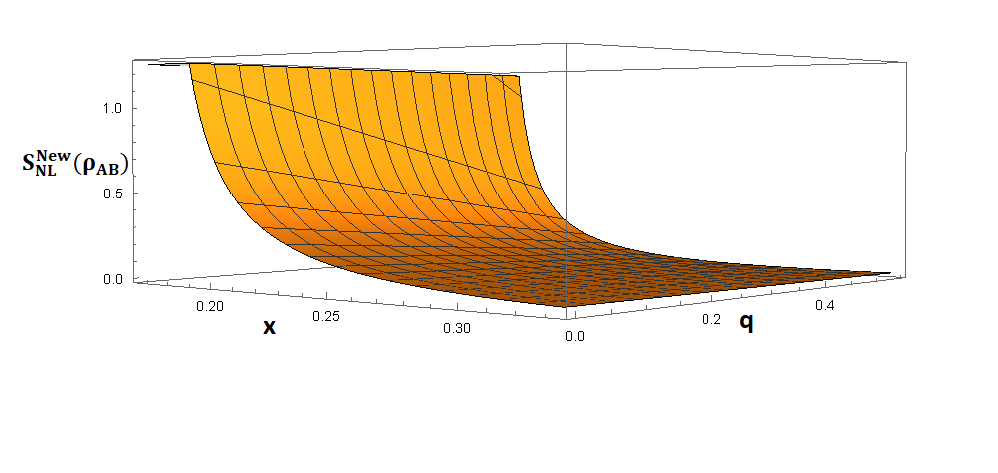}
		\caption{The curve represents the non-zero value of $S_{NL}^{New}(\rho_{AB})$ for the state $\rho_{AB}$. Here, $x$ denotes the state parameter and $q$ lies in the range  (0,0.55). }
	\end{figure}
	\subsection{Relation between $S_{NL}({\rho^{ent}_{AB}})$ and the quantity $M(\rho^{ent}_{AB})$}
	In this subsection, we consider a two-qubit entangled state described by the density operator $\rho^{ent}_{AB}$, we obtain the relationship between the strength of the non-locality $S_{NL}(\rho^{ent}_{AB})$ and the quantity $M(\rho^{ent}_{AB})$. To derive the required relationship, we need a few lemmas which are given below:\\
	\textbf{Lemma-1:} If $P^{max}(\rho^{ent}_{AB})$ denotes the maximum probability of winning the game via the shared state $\rho^{ent}_{AB}$ between the two players, then the upper bound of $P^{max}(\rho^{ent}_{AB})$ in terms of $M(\rho^{ent}_{AB})$ is given by
	\begin{eqnarray}
		P^{max}(\rho^{ent}_{AB})\leq \frac{1}{2}(\frac{\sqrt{M(\rho^{ent}_{AB})}}{2}+1)
		\label{L1}
	\end{eqnarray}
	\textbf{Proof:} Recalling (\ref{pmax}), $P^{max}(\rho^{ent}_{AB})$ can be re-written as
	\begin{eqnarray}
		P^{max}(\rho^{ent}_{AB})=\frac{1}{2}(1+\frac{ \langle B_{CHSH}\rangle_{\rho^{ent}_{AB}}}{4})
		\label{L11}
	\end{eqnarray}
	Let us denote $\langle B_{max} \rangle_{\rho^{ent}_{AB}}=max_{B_{CHSH}}\langle B_{CHSH} \rangle_{\rho^{ent}_{AB}}$. Therefore, $P^{max}(\rho^{ent}_{AB})$ given in (\ref{L11}) reduces to
	\begin{eqnarray}
		P^{max}(\rho^{ent}_{AB})&\leq& \frac{1}{2}(1+\frac{\langle B_{max} \rangle_{\rho^{ent}_{AB}}}{4})\nonumber\\&=& \frac{1}{2}(1+\frac{  \sqrt{M(\rho^{ent}_{AB})}}{2})
		\label{L12}
	\end{eqnarray}
	In the last line, we have used $\langle B_{max} \rangle_{\rho^{ent}_{AB}}=2\sqrt{M(\rho^{ent}_{AB})}$ \cite{horo3}. 
	%The lower bound of $P^{max}(\rho_{ent})$ is clear from the fact that $M(\rho_{ent})>1$ for $\rho_{ent}$ violate the CHSH inequality. 
	Hence proved.\\
	Using Lemma-1 and upper bound of $M(\rho^{ent}_{AB})$ i.e., $M(\rho^{ent}_{AB})\leq 2$, it can be easily observed that $P^{max}(\rho^{ent}_{AB})\leq \frac{1}{2}(1+\frac{1}{\sqrt{2}})$.\\
	\textbf{Lemma-2:} If $W_{CHSH}$ denotes the witness operator detecting the two-qubit entangled state $\rho^{ent}_{AB}$, then the lower bound of $M(\rho^{ent}_{AB})$ is given by
	\begin{eqnarray}
		M(\rho^{ent}_{AB})\geq [1-\frac{1}{2}{Tr\rm}[W_{CHSH}\rho^{ent}_{AB}]]^{2}
		\label{L2}
	\end{eqnarray} 
	\textbf{Proof:} From (\ref{pmax}), $\langle B_{CHSH}\rangle_{\rho^{ent}_{AB}}$ can be expressed as
	\begin{eqnarray}
		\langle B_{CHSH}\rangle_{\rho^{ent}_{AB}}=8P^{max}(\rho^{ent}_{AB})-4
		\label{L21}
	\end{eqnarray}
	Using $\langle B_{CHSH}\rangle_{\rho^{ent}_{AB}}\leq\langle B_{max}\rangle_{\rho^{ent}_{AB}}$, the equation can be re-expressed as
	\begin{eqnarray}
		8P^{max}(\rho^{ent}_{AB})-4\leq \langle B_{max}\rangle_{\rho^{ent}_{AB}}=2\sqrt{M(\rho^{ent}_{AB})}
		\label{L22}
	\end{eqnarray}
	Using Result-1 and simplifying (\ref{L22}), we get the required result. Hence proved.\\
	Now, we are in a position to connect $S_{NL}(\rho^{ent}_{AB})$ and $M(\rho^{ent}_{AB})$.\\
	\textbf{Result-3}:- If $\rho^{ent}_{AB}$ denotes any two-qubit entangled state, which violates the CHSH inequality and is detected by $W_{CHSH}$, then
	\begin{eqnarray}
		S_{NL}(\rho^{ent}_{AB})< \frac{\sqrt{M(\rho^{ent}_{AB})}-1}{4}
		\label{r61}		
	\end{eqnarray} 
	\textbf{Proof:} Since the CHSH witness operator $W_{CHSH}$ detects the entangled state $\rho^{ent}_{AB}$, so $S_{NL}(\rho^{ent}_{AB})$
	is given by
	\begin{eqnarray}
		S_{NL}(\rho^{ent}_{AB})=-\frac{{Tr\rm}[W_{CHSH}\rho^{ent}_{AB}]}{8} 
		\label{pr61}		
	\end{eqnarray}
	Using Lemma-2, $S_{NL}(\rho^{ent}_{AB})$ can be re-expressed in terms of $M(\rho^{ent}_{AB})$ as
	\begin{eqnarray}
		S_{NL}(\rho^{ent}_{AB})< \frac{\sqrt{M(\rho^{ent}_{AB})}-1}{4}
		\label{pr62}		
	\end{eqnarray}
	Hence Proved.\\
	Using Result-3, and the fact $M(\rho^{ent}_{AB})\leq 2$, we get the upper bound of $S_{NL}(\rho^{ent}_{AB})$, which is given by
	\begin{eqnarray}
		S_{NL}(\rho^{ent}_{AB})< \frac{\sqrt{2}-1}{4}
		\label{pr63}		
	\end{eqnarray}
	So far, we have discussed the relationship between $S_{NL}(\rho^{ent}_{AB})$ and $M(\rho^{ent}_{AB})$ when $M(\rho^{ent}_{AB})> 1$. But what if, $M(\rho^{ent}_{AB})\leq 1$? Let us now discuss this case in the form of another result that can be stated as:\\
	\textbf{Result-4:} If we suppose that the two-qubit entangled state $\rho^{ent}_{AB}$ satisfies the CHSH inequality i.e., $M(\rho^{ent}_{AB})\leq 1$, and further, if it is not detected by the witness operator $W_{CHSH}$, then the relation between $S_{NL}(\rho^{ent}_{AB})$ and $M(\rho^{ent}_{AB})$ is given by
	\begin{eqnarray}
		0 < S_{NL}(\rho^{ent}_{AB}) \leq q (\frac{\sqrt{M(\rho^{ent}_{AB})}-1}{4})+(1-q)K
		\label{snlmnd}		
	\end{eqnarray}
	where $K=\frac{{Tr\rm}[W_{CHSH}\rho_{AB}^{ent}(\rho_{AB}^{ent})^{T_{B}}]}{4N(\rho_{AB}^{ent})}$ and $q$ satisfy the inequality
	\begin{eqnarray}
		0 \leq q < \frac{4K}{1-\sqrt{M(\rho^{ent}_{AB})}+4K}
		\label{rangeq}		
	\end{eqnarray}
	\textbf{Proof:} If the two-qubit entangled state $\rho^{ent}_{AB}$ is not detected by the witness operator $W_{CHSH}$
	then $P^{max}\leq \frac{3}{4}$. Thus, the strength of the non-locality $S^{New}_{NL}(\rho^{ent}_{AB})$ of $\rho^{ent}_{AB}$ may be defined by (\ref{snldef}). Therefore recalling (\ref{snldef}), we get
	\begin{eqnarray}
		S_{NL}(\rho^{ent}_{AB})&=& q(P^{max}-\frac{3}{4})+(1-q)K \nonumber\\&\leq& q(\frac{\sqrt{M(\rho^{ent}_{AB})}-1}{4})+(1-q)K 
		\label{snldef1}
	\end{eqnarray}
	In the second line, we have used inequality (\ref{L1}).\\
	Since the inequality (\ref{snldef1}) gives the upper bound of $S_{NL}(\rho^{ent}_{AB})$ in terms of $M(\rho^{ent}_{AB})$, so it may happen that the value of $S_{NL}(\rho^{ent}_{AB})$ may be negative also, which is not acceptable. Thus, to make it positive, we have to put some restrictions on $q$. Therefore, We can choose q in such a way that the inequality (\ref{qub}) holds. The inequality (\ref{qub}) may be re-expressed in terms of $M(\rho^{ent}_{AB})$ as
	\begin{eqnarray}
		0 \leq q < \frac{4K}{1-\sqrt{M(\rho^{ent}_{AB})}+4K}
		\label{qub1}
	\end{eqnarray}    
	Hence proved.\\
	Further, employing the condition $M(\rho^{ent}_{AB})\leq 1$ again, it can be easily shown that the inequality (\ref{snldef1}) reduces to
	\begin{eqnarray}
		S^{New}_{NL}(\rho^{ent}_{AB})\leq (1-q)K 
		\label{snldef2}
	\end{eqnarray}
	Hence, we have shown here that we are capable of detecting the non-locality of $\rho^{ent}_{AB}$ even if $M(\rho^{ent}_{AB})\leq 1$, for some entangled state $\rho^{ent}_{AB}$.
	\section{Strength of the non-locality of two-qubit entangled system determined by optimal witness operator}  
	In the previous section, we found that, there may exist a shared entangled state $\rho^{ent}_{AB}$ which is not detected by witness operator $W_{CHSH}$, and as a consequence, the maximum probability $P^{max}$ of winning the game played between two distant players with $\rho^{ent}_{AB}$ must be less than or equal to $\frac{3}{4}$. But just by merely observing this fact, we cannot say that the strength of the non-locality of the state $\rho^{ent}_{AB}$ is zero as there exist other witness operators that may detect it. But the problem is that there does not exist any general relationship between $P^{max}$ and any witness operator $W^{a}$ different from $W_{CHSH}$. Thus, in this perspective, we can ask the following question: for any two-qubit entangled state $\rho^{ent}_{AB}$ shared between two distant players playing the XOR game and if, ${Tr\rm}(W_{CHSH}\rho^{ent}_{AB})\geq 0$ and ${Tr\rm}(W^{a}\rho^{ent}_{AB})< 0$,  then can we measure the strength of the non-locality of two qubit-entangled state $\rho^{ent}_{AB}$? We investigate this question for a particular case, $W^{a}=W^{opt}$, and $W^{opt}$ denotes the optimal witness operator. The reason behind this choice is that the optimal witness operator detects the maximum number of entangled states.
	\subsection{Derivation of witness operator inequality}
	In this subsection, we start with the derivation of witness operator inequality using Bell-CHSH inequality. To achieve this inequality, we may consider the optimal witness operator as $W^{opt}=(|\psi\rangle_{AB}\langle \psi|)^{T_{B}}$, where $|\psi\rangle_{AB}=\frac{1}{\sqrt{2}}(|00\rangle+|11\rangle)$ and $T_{B}$ denote the partial transposition with respect to subsystem $B$. In the second step, we establish a relationship between the optimal witness operator $W^{opt}$ and the CHSH witness operator $W_{CHSH}$, and then we derive the lower and upper bound of  $W_{CHSH}^{xy}+W_{CHSH}^{xz}+W_{CHSH}^{yz}$, when the optimal witness operator $W^{opt}$ detects the entangled state $\rho^{ent}_{AB}$.\\ %Lastly, we derive the upper bound of the strength of the non-locality $S_{NL}$ of the state under investigation.\\
	To start with, let us consider $W^{opt}$ that may be expressed in terms of the Bell operators $B_{xy}$, $B_{xz}$, and $B_{yz}$ as \cite{hyllus}
	\begin{equation}
		W^{opt}= \frac{1}{4}[I_{4}+\frac{1}{2\sqrt{2}}(B_{xy}+B_{xz}+B_{yz})]
		\label{wit1}
	\end{equation} 
	The expectation value of $W^{opt}$ with respect to the two-qubit density operator $\rho^{ent}_{AB}$ is given by 
	\begin{equation}
		{Tr\rm}[W^{opt}\rho^{ent}_{AB}]= \frac{1}{4}[1+\frac{1}{2\sqrt{2}}(\langle B_{xy}\rangle_{\rho^{ent}_{AB}}+
		\langle B_{xz}\rangle_{\rho^{ent}_{AB}}\nonumber\\+\langle B_{yz}\rangle_{\rho^{ent}_{AB}})]
		\label{b122}
	\end{equation} 
	Recalling (\ref{R1}) and adding the expression of $P_{ij}^{max}$ for different $i$ $\&$ $j$, we get 
	\begin{eqnarray}
		\sum_{\substack{i,j=x,y,z \\ i\neq j}}P_{i,j}^{max}=\frac{3}{2}+\frac{\sum_{\substack{i,j=x,y,z \\ i\neq j}}\langle B_{ij}\rangle_{\rho^{ent}_{AB}}}{8} 
		\label{E1}
	\end{eqnarray}
	Using (\ref{wit1}), we can re-express (\ref{E1}) in terms of the expectation value of $W^{opt}$ as
	\begin{equation}
		P_{xy}^{max}+ P_{yz}^{max}+ P_{zx}^{max}= \frac{3}{2}-\frac{1}{2\sqrt{2}}+\sqrt{2}{Tr\rm}[W^{opt}\rho^{ent}_{AB}]
		\label{E2}
	\end{equation}
	We should note an important fact that the expectation value of CHSH witness operator $W_{CHSH}$ is positive i.e., $\langle W_{CHSH}\rangle \geq 0$ when $\langle B_{CHSH}\rangle$ lying in the subinterval $[-2\sqrt{2},0]$, while it is positive or negative according to $\langle B_{CHSH}\rangle \in (0,2]$ or $\langle B_{CHSH}\rangle \in (2,2\sqrt{2}]$. Since, we assume that the state $\rho^{ent}_{AB}$ satisfies the Bell's inequality in every setting, so we consider $-2\leq \langle B_{ij}\rangle_{\rho^{ent}_{AB}} \leq 2,~~i,j=x,y,z;~ i \neq j$.  Thus, using (\ref{R1}) in the interval $[-2,2]$, we get 
	\begin{eqnarray}
		\frac{1}{4} \leq P_{ij}^{max}\leq \frac{3}{4},~~\forall i,j=x,y,z~~ \&~~ i \neq j
		\label{E3}
	\end{eqnarray}
	Therefore, using (\ref{E3}) in (\ref{E2}) and after simplifying it, we get
	\begin{equation}
		-0.28033 \leq {Tr\rm}[W^{opt}\rho^{ent}_{AB}] \leq 0.78033
		\label{E422} 
	\end{equation} 
	Since the inequality (\ref{E422}) is derived using the Bell-CHSH inequality, and it involves the expectation value of the witness operator, so it may be termed as witness operator inequality. This inequality clearly shows that there exists a witness operator such as $W^{opt}$ that may detect the entangled state $\rho^{ent}_{AB}$, which may not be identified by the Bell operator $B_{ij}~~(i,j=x,y,z; ~ i \neq j)$.
	The existence of the subinterval $[-0.28033, 0]$ of the witness operator inequality indicates the fact that we may have entangled states $\rho^{ent}_{AB}$ that can be detected by $W^{opt}$, although it satisfies the Bell-CHSH inequality.\\
	Now we are in a position to derive the lower and upper bound of $W_{CHSH}^{xy}+W_{CHSH}^{xz}+W_{CHSH}^{yz}$. To derive the required lower and upper bound, we are exploiting the subinterval $[-0.28033, 0]$, where $W^{opt}$ detects the entangled state $\rho^{ent}_{AB}$. We should note here a crucial point that the state $\rho^{ent}_{AB}$ is not detected by any of the operators $W_{CHSH}^{xy}$, $W_{CHSH}^{xz}$, and $W_{CHSH}^{yz}$.\\
	\textbf{Result-5:} If $\rho^{ent}_{AB}$ denotes an entangled state which is not detected by $W_{CHSH}^{xy}$, $W_{CHSH}^{yz}$, and $W_{CHSH}^{zx}$, and $W^{opt}$ is an optimal witness operator such that ${Tr\rm}[W^{opt}\rho^{ent}_{AB}]\in [-0.28033,0]$, then 
	\begin{eqnarray}
		8.82843 &\leq& \langle W_{CHSH}^{xy}\rangle_{\rho^{ent}_{AB}}+\langle W_{CHSH}^{yz}\rangle_{\rho^{ent}_{AB}}\nonumber\\&+&\langle W_{CHSH}^{xz}\rangle_{\rho^{ent}_{AB}} \leq 11.9997
		\label{result-3}
	\end{eqnarray}
	\textbf{Proof:} To start the derivation of the bounds, let us first express the expectation value of $W^{opt}$ in terms of the expectation value of $W_{CHSH}^{xy}$, $W_{CHSH}^{xz}$ and $W_{CHSH}^{yz}$. It is given by
	\begin{eqnarray}
		{Tr\rm}[W^{opt}\rho^{ent}_{AB}]&=&\frac{1}{4}[1+\frac{1}{2\sqrt{2}}(\langle B_{xy}\rangle_{\rho^{ent}_{AB}}+\langle B_{yz}\rangle_{\rho^{ent}_{AB}}\nonumber\\ &+&\langle B_{xz}\rangle_{\rho^{ent}_{AB}})] \nonumber\\
		&=&\frac{1}{4}[1+\frac{1}{2\sqrt{2}}(6-\langle W_{CHSH}^{xy}\rangle_{\rho^{ent}_{AB}}\nonumber\\&-&\langle W_{CHSH}^{yz}\rangle_{\rho^{ent}_{AB}} -\langle W_{CHSH}^{xz}\rangle_{\rho^{ent}_{AB}})]
		\label{eq30}
	\end{eqnarray}
	Considering the witness operator inequality in the negative subinterval, i.e. when ${Tr\rm}[W^{opt}\rho^{ent}_{AB}] \in [-0.2803,0]$, (\ref{eq30}) reduces to the inequality 
	\begin{eqnarray}
		8.82843 &\leq& \langle W_{CHSH}^{xy}\rangle_{\rho^{ent}_{AB}}+\langle W_{CHSH}^{yz}\rangle_{\rho^{ent}_{AB}}\nonumber\\&+&\langle W_{CHSH}^{xz}\rangle_{\rho^{ent}_{AB}} \leq 11.9997
	\end{eqnarray}
	Hence proved.\\
	Thus, the witness operator inequality (\ref{result-3}) in the negative region gives the lower and upper bound of $\langle W_{CHSH}^{xy}\rangle_{\rho^{ent}_{AB}}+\langle W_{CHSH}^{yz}\rangle_{\rho^{ent}_{AB}}+\langle W_{CHSH}^{xz}\rangle_{\rho^{ent}_{AB}}$, provided $\langle W_{CHSH}^{xy}\rangle_{\rho^{ent}_{AB}}\geq 0,\langle W_{CHSH}^{yz}\rangle_{\rho^{ent}_{AB}}\geq 0,\langle W_{CHSH}^{zx}\rangle_{\rho^{ent}_{AB}}\geq 0$.\\
	To illustrate our result, let us consider the state described by the density operator $\rho_{AB}^{(3)}$ 
	\begin{eqnarray}
		\rho_{AB}^{(3)}&=& \frac{1}{4}[I\otimes I-0.01\sigma_{1}\otimes I +0.002 I \otimes \sigma_{3}-0.7\sigma_{1} \otimes \sigma_{1} \nonumber\\&-&0.7\sigma_{2} \otimes \sigma_{2}-0.67 \sigma_{3} \otimes \sigma_{3}] 
	\end{eqnarray} 
	We find that the state $\rho_{AB}^{(3)}$ is an entangled state, but it satisfies the Bell-CHSH inequality in different settings as $\langle B_{xy}\rangle_{\rho_{AB}^{(3)}}=-1.9799$, $\langle B_{yz}\rangle_{\rho_{AB}^{(3)}}=-1.93747$, $\langle B_{xz}\rangle_{\rho_{AB}^{(3)}}=-1.93747$. Further, we find that the state $\rho_{AB}^{(3)}$ is not detected by the CHSH witness operator as $\langle W^{xy}_{CHSH}\rangle_{\rho_{AB}^{(3)}}=3.9799\geq 0$,$\langle W^{xz}_{CHSH}\rangle_{\rho_{AB}^{(3)}}=3.93747\geq 0$ and $\langle W^{yz}_{CHSH}\rangle_{\rho_{AB}^{(3)}}=3.93747\geq 0$. \\
	Let us now probe whether the state $\rho_{AB}^{(3)}$ is detected by $W^{opt}$ or not. To investigate this, let us calculate the expectation value of $W^{opt}$ with respect to the state $\rho_{AB}^{(3)}$ as
	\begin{eqnarray}
		{Tr\rm}[W^{opt}\rho_{AB}^{(3)}]&=& \frac{1}{4}[1+\frac{1}{2\sqrt{2}}(\langle B_{xy}\rangle_{\rho_{AB}^{(3)}}+\langle B_{xz}\rangle_{\rho_{AB}^{(3)}}\nonumber\\&+&\langle B_{yz}\rangle_{\rho_{AB}^{(3)}})]\nonumber\\&=&-0.2675
	\end{eqnarray} 
	Therefore, ${Tr\rm}[W^{opt}\rho_{AB}^{(3)}]$ satisfies the witness operator inequality, and thus, one can easily verify that 
	$ \langle W^{xy}_{CHSH}\rangle_{\rho_{AB}^{(3)}}+\langle W^{yz}_{CHSH}\rangle_{\rho_{AB}^{(3)}}+\langle W^{xz}_{CHSH}\rangle_{\rho_{AB}^{(3)}}=11.85484$ satisfy the inequality (\ref{result-3}).
	\subsection{Upper bound for the strength of non-locality of two-qubit entangled system detected by optimal witness operator}
	In this subsection, we first derive the inequality that provides the upper bound of the maximum probability of winning in terms of the expectation value of $W^{opt}$. Therefore, we have established the connection between the maximum probability of winning and the expectation value of $W^{opt}$. This connection enables us to estimate the strength of the non-locality of an entangled state which is undetected by $W_{CHSH}$ but detected by $W^{opt}$. The following result educates us about the question that we have in the starting paragraph of this section.\\
	\textbf{Result-6:-} If the quantum state $\rho^{ent}_{AB}$ satisfies the Bell-CHSH inequality in $xy-$, $yz-$ and $zx-$ setting i.e. if $-2\leq \langle B_{ij}\rangle_{\rho^{ent}_{AB}}\leq 2,~\forall~i,j=x,y,z;~ i\neq j$, and if the state $\rho^{ent}_{AB}$ may be identified as an entangled state by the witness operator $W^{opt}$ given in (\ref{wit1}), then the strength of the non-locality of $\rho^{ent}_{AB}$ may be estimated by the inequality
	\begin{eqnarray} 
		S_{NL}(\rho^{ent}_{AB})&\leq& \frac{3}{4}-\frac{1}{2\sqrt{2}}+\sqrt{2}{Tr\rm}[W^{opt}\rho^{ent}_{AB}]
		\label{exp23}
	\end{eqnarray} 
	\textbf{Proof:} Let us assume $max\{P_{xy}^{max}, P_{xz}^{max}, P_{yz}^{max}\}=P_{xy}^{max}$. Then we can have the following inequality
	\begin{equation}
		P_{xy}^{max}\leq P_{xy}^{max}+ P_{yz}^{max}+ P_{zx}^{max}
		\label{ineq1}
	\end{equation}
	Recalling the expression given in (\ref{E2}) and using (\ref{ineq1}), we get
	\begin{eqnarray}
		&&P_{xy}^{max}\leq \frac{3}{2}-\frac{1}{2\sqrt{2}}+\sqrt{2}{Tr\rm}[W^{opt}\rho^{ent}_{AB}]
		\nonumber\\&& \Rightarrow P_{xy}^{max}-\frac{3}{4}\leq  U
		\label{ineq2}
	\end{eqnarray}
	where $U=\frac{3}{4}-\frac{1}{2\sqrt{2}}+\sqrt{2}{Tr\rm}[W^{opt}\rho^{ent}_{AB}]$.\\
	Our task is now to check whether the upper bound of $P^{max}_{xy}-\frac{3}{4}$ is positive when $W^{opt}$ detects the entangled state $\rho^{ent}_{AB}$. We have to check this because this verification will indicate that there is a possibility of detecting non-locality via $W^{opt}$. The truthfulness of the above statement is given in Table-I:\\
	\begin{table*}[htbp!]\centering
		\begin{tabular} {|c|c|c|}\hline 
			%\multicolumn{7}{c}{text}\hline 
			S.No.& ${Tr\rm}[W^{opt}{\rho^{ent}_{AB}}]$ & U\\ \hline
			1 & 0 & 0.39645 \\ \hline 
			2 & -0.05 & 0.325736\\ \hline 
			3  & -0.10 & 0.255025\\ \hline 
			4 & -0.15 & 0.184315\\ \hline 
			5 & -0.20 & 0.113604\\ \hline 
			6& -0.25 & 0.0428932\\ \hline 
			7 & -0.28033 & 0.0001 \\ \hline 					
		\end{tabular}
		\caption{The table shows the upper bound of $S_{NL}(\rho^{ent}_{AB})$ when ${Tr\rm}[W^{opt}\rho^{ent}_{AB}]\in [-0.28033,0]$}
	\end{table*}
	We are now in a position to estimate the non-locality of the entangled state described by the density operator $\rho^{ent}_{AB}$. Therefore, using the definition of the strength of the non-locality $S_{NL}^{(xy)}(\rho^{ent}_{AB})$ given in (\ref{snl23}), the inequality (\ref{ineq2}) reduces to 
	\begin{eqnarray}
		S_{NL}^{(xy)}(\rho^{ent}_{AB}) \leq \frac{3}{4}-\frac{1}{2\sqrt{2}}+\sqrt{2}{Tr\rm}[W^{opt}\rho^{ent}_{AB}]
		\label{ineq6}
	\end{eqnarray}   
	Similarly, if we assume either $max\{P_{xy}^{max}, P_{xz}^{max}, P_{yz}^{max}\}=P_{xz}^{max}$ or $max\{P_{xy}^{max}, P_{xz}^{max}, P_{yz}^{max}\}=P_{yz}^{max}$ then, we get the same result. Since the upper bound of the strength of the non-locality does not depend on any particular setting, so the inequality (\ref{ineq6}) may be re-expressed as
	\begin{eqnarray}
		S_{NL}(\rho^{ent}_{AB}) \leq \frac{3}{4}-\frac{1}{2\sqrt{2}}+\sqrt{2}{Tr\rm}[W^{opt}\rho^{ent}_{AB}]
		\label{ineq61}
	\end{eqnarray} 
	Hence the theorem is proved.\\
	To illustrate our result, let us consider the state described by the density operator $\rho_{n}$, which is given by
	\begin{eqnarray}
		\rho_{n}=\begin{pmatrix} 
			\frac{1-a}{6} & 0& 0& 0.0005\\
			0 & \frac{5}{6}-a & -0.251 & 0\\
			0 & -0.251 & a & 0\\
			0.0005 & 0& 0& \frac{a}{6} \\
		\end{pmatrix},~\frac{1}{10}<a<\frac{13}{20}\nonumber\\
	\end{eqnarray}
	Applying the partial transposition criterion, we can say that the state $\rho_{n}$ is an entangled state. The state satisfies the Bell-CHSH inequality, as we find that $\langle B_{xy}\rangle_{\rho_{n}}=-1.41987$, $\langle B_{yz}\rangle_{\rho_{n}}=-1.65416$, \text{and} $\langle B_{xz}\rangle_{\rho_{n}}=-1.65133$. But, the state $\rho_{n}$ is detected by $W^{opt}$ as ${Tr\rm}[W^{opt}\rho_{n}]=-0.167667<0$. Although the state $\rho_{n}$ satisfies the Bell-CHSH inequality in different settings, but it is detected by $W^{opt}$. Thus, we can use our Result-6, for the estimation of the non-locality of $\rho_{n}$. Therefore, the strength of the non-locality of the entangled state $\rho_{n}$ is given by
	\begin{eqnarray}
		S_{NL}(\rho_{n}) \leq 0.15933
		\label{ineq62}
	\end{eqnarray} 
	\section{Expression for the strength of the non-locality of two-qubit entangled state in terms of measurement parameter and state parameter}
	\textbf{Theorem-1:-} If Alice (A) and Bob (B) share any arbitrary two-qubit entangled state described by the density operator $\rho_{AB}^{ent}$ given in (\ref{gen2qbitst22}), 
	and if the maximized winning probability $P^{max}$ satisfies $P^{max} \leq \frac{3}{4}$ then 
	\begin{eqnarray}
		&&	c_{1}[\lambda_{1}^{(0)}(\mu_{1}^{(0)}-\mu_{1}^{(1)})+\lambda_{1}^{(1)}(\mu_{1}^{(0)}+\mu_{1}^{(1)})]\nonumber \\
		&+& c_{2}[\lambda_{2}^{(0)}(\mu_{2}^{(0)}-\mu_{2}^{(1)}) +\lambda_{2}^{(1)}(\mu_{2}^{(0)}+\mu_{2}^{(1)})] \nonumber\\
		&+& c_{3}[\lambda_{3}^{(0)}(\mu_{3}^{(0)}-\mu_{3}^{(1)})+\lambda_{3}^{(1)}(\mu_{3}^{(0)}+\mu_{3}^{(1)})]\leq 2. 
	\end{eqnarray}
	where $\lambda_{j}^{i}\in \mathbb{R}^{3}$ and $\mu_{j}^{i}\in \mathbb{R}^{3}$ $(i=0,1;j=1,2)$ denote the real parameter of the Bell operator, which satisfies
	\begin{eqnarray}
		(\lambda_{1}^{(i)})^{2}+(\lambda_{2}^{(i)})^{2}+(\lambda_{3}^{(i)})^{2} =1,~~i=0,1 \\
		(\mu_{1}^{(i)})^{2}+(\mu_{2}^{(i)})^{2}+(\mu_{3}^{(i)})^{2} =1,~~i=0,1 
	\end{eqnarray}
	\textbf{Proof:-} Let us start with the Bell-CHSH operator $B_{CHSH}$, which is given by 
	\begin{eqnarray*}
		B_{CHSH}= A_{0}\otimes B_{0}-A_{0}\otimes B_{1}+A_{1}\otimes B_{0}+A_{1}\otimes B_{1}
	\end{eqnarray*}
	The witness operator $W_{CHSH}$ can be constructed from the Bell-CHSH operator as
	\begin{equation}
		W_{CHSH}=2I \otimes I-A_{0}\otimes B_{0}+A_{0}\otimes B_{1}-A_{1}\otimes B_{0}-A_{1}\otimes B_{1}
		\label{2}
	\end{equation}
	where the Hermitian operators $A_{0},A_{1},B_{0},B_{1}$ can be expressed in terms of the Pauli matrices $\sigma_{i},~i=x,y,z$ as
	\begin{eqnarray}
		A_{0}&=&\lambda_{1}^{0}\sigma_{x}+\lambda_{2}^{0}\sigma_{y}+\lambda_{3}^{0}\sigma_{z}\nonumber \\
		A_{1}&=&\lambda_{1}^{1}\sigma_{x}+\lambda_{2}^{1}\sigma_{y}+\lambda_{3}^{1}\sigma_{z}\nonumber \\ 
		B_{0}&=&\mu_{1}^{0}\sigma_{x}+\mu_{2}^{0}\sigma_{y}+\mu_{3}^{0}\sigma_{z}\nonumber \\
		B_{1}&=&\mu_{1}^{1}\sigma_{x}+\mu_{2}^{1}\sigma_{y}+\mu_{3}^{1}\sigma_{z}
		\label{hermop}
	\end{eqnarray}
	Recalling the two-qubit state $\rho_{AB}^{ent}$ given in (\ref{gen2qbitst22}), and then let us calculate the expectation value of $W_{CHSH}$ with respect to the state $\rho_{AB}^{ent}$. The expectation value is given by
	\begin{eqnarray}
		{Tr\rm}[W_{CHSH}\rho_{AB}^{ent}]= 2-\frac{1}{4}[&&\sum_{j=x,y,z}c_{j}\{{Tr\rm}(A_{0}\sigma_{j})\nonumber\\&\times&  {Tr\rm}[(B_{0}-B_{1})\sigma_{j}] + {Tr\rm}(A_{1}\sigma_{j})\nonumber \\ &\times&{Tr\rm}[(B_{0}+B_{1})\sigma_{j}]\}]
		\label{exp122}
	\end{eqnarray}
	Using (\ref{hermop}) in (\ref{exp122}), we get 
	\begin{eqnarray}
		{Tr\rm}[W_{CHSH}\rho_{AB}^{ent}]&=& 2- 	\{c_{1}[\lambda_{1}^{(0)}(\mu_{1}^{(0)}-\mu_{1}^{(1)})+\lambda_{1}^{(1)}(\mu_{1}^{(0)}\nonumber\\&&+\mu_{1}^{(1)})] +c_{2}[\lambda_{2}^{(0)}(\mu_{2}^{(0)}-\mu_{2}^{(1)}) 
		+\lambda_{2}^{(1)}\times\nonumber\\&&(\mu_{2}^{(0)}+\mu_{2}^{(1)})] +c_{3}[\lambda_{3}^{(0)}(\mu_{3}^{(0)}-\mu_{3}^{(1)})+
		\nonumber\\&&\lambda_{3}^{(1)}(\mu_{3}^{(0)}+\mu_{3}^{(1)})]\}
	\end{eqnarray}
	From Result-1, it is clear that $P^{max}\leq \frac{3}{4}$, only when ${Tr\rm}[W_{CHSH}\rho_{AB}^{ent}]\geq 0$. Therefore, 
	\begin{eqnarray}
		&& {Tr\rm}[W_{CHSH}\rho_{AB}^{ent}]\geq 0 \implies \nonumber \\ && c_{1}[\lambda_{1}^{(0)}(\mu_{1}^{(0)}-\mu_{1}^{(1)})+\lambda_{1}^{(1)}(\mu_{1}^{(0)}+\mu_{1}^{(1)})]\nonumber \\
		&+&c_{2}[\lambda_{2}^{(0)}(\mu_{2}^{(0)}-\mu_{2}^{(1)})+\lambda_{2}^{(1)}(\mu_{2}^{(0)}+\mu_{2}^{(1)})] \nonumber \\
		&+&c_{3}[\lambda_{3}^{(0)}(\mu_{3}^{(0)}-\mu_{3}^{(1)})+\lambda_{3}^{(1)}(\mu_{3}^{(0)}+\mu_{3}^{(1)})]\leq 2. \nonumber
	\end{eqnarray}
	Hence Proved.\\ 
	\textbf{Corollary-1:-} If the following inequality is satisfied by any two-qubit arbitrary entangled state $\rho_{AB}^{ent}$, 
	\begin{eqnarray}
		&&c_{1}[\lambda_{1}^{(0)}(\mu_{1}^{(0)}-\mu_{1}^{(1)})+\lambda_{1}^{(1)}(\mu_{1}^{(0)}+\mu_{1}^{(1)})]  \nonumber \\
		&+&c_{2}[\lambda_{2}^{(0)}(\mu_{2}^{(0)}-\mu_{2}^{(1)}) +\lambda_{2}^{(1)}(\mu_{2}^{(0)}+\mu_{2}^{(1)})]\nonumber \\
		&+& 	c_{3}[\lambda_{3}^{(0)}(\mu_{3}^{(0)}-\mu_{3}^{(1)})+\lambda_{3}^{(1)}(\mu_{3}^{(0)}+\mu_{3}^{(1)})] >2 
	\end{eqnarray}
	and the state is detected by $W_{CHSH}$, then $P^{max} >\frac{3}{4}$.\\
	\textbf{Proof:} This corollary follows from Result-1.\\
	Now we are in a position to measure the strength of the non-locality of any general two-qubit entangled state. The expression of the strength of the non-locality can be expressed in terms of the measurement parameters and state parameters, and it is given in the result below:\\
	\textbf{Result-7:} If any arbitrary two-qubit state described by the density operator $\rho_{AB}^{ent}$ given in (\ref{gen2qbitst22}) represents an entangled state, which is detected by the witness operator $W_{CHSH}$ then its non-locality can be determined using the following formula
	\begin{eqnarray}
		S_{NL}(\rho_{AB}^{ent})&=&\frac{1}{8}[c_{1}(\lambda_{1}^{(0)}(\mu_{1}^{(0)}-\mu_{1}^{(1)})+\lambda_{1}^{(1)}(\mu_{1}^{(0)}+\mu_{1}^{(1)}))  \nonumber \\
		&+&c_{2}(\lambda_{2}^{(0)}(\mu_{2}^{(0)}-\mu_{2}^{(1)}) +\lambda_{2}^{(1)}(\mu_{2}^{(0)}+\mu_{2}^{(1)}))\nonumber \\
		&+& 	c_{3}(\lambda_{3}^{(0)}(\mu_{3}^{(0)}-\mu_{3}^{(1)})+\lambda_{3}^{(1)}(\mu_{3}^{(0)}+\mu_{3}^{(1)}))\nonumber\\&-&2] 
	\end{eqnarray}
	\section{Applications}
	In this section, we will discuss two applications of the introduced quantity $S_{NL}(\rho^{ent}_{AB})$ such as (i) application of $S_{NL}(\rho^{ent}_{AB})$ in the determination of the genuine non-locality of two particular classes of three-qubit GHZ-state and W-state, and (ii) application of $S_{NL}(\rho^{ent}_{AB})$ in finding the upper limit of the power of the controller in controlled quantum teleportation.   
	\subsection{Linkage between the strength of the non-locality of two-qubit entangled state and the expectation value of the Svetlichny operator with respect to a pure three-qubit state}
	In this section, we give a brief discussion about the non-locality of the three-qubit state, and then we establish a relationship between the two-qubit non-locality with the non-locality of the pure three-qubit state. We measure the strength of the two-qubit non-locality by $S_{NL}$, and the pure three-qubit non-locality is measured by the expectation value of the Svetlichny operator.\\ 
	Let us consider a tripartite system describing a pure three-qubit state. In a three-qubit state, there may exist different types of correlation. The correlation may exist either between two qubits only or between all three qubits. The correlations are genuinely tripartite non-local if the correlations cannot be simulated by a hybrid (non-local)-local ensemble of a three-qubit system. Here, a hybrid (non-local)-local ensemble of a three-qubit system means that any two qubits are non-locally correlated but it is locally correlated, with the third qubit. The genuine tripartite non-local correlation exists in the three-qubit state $\rho_{ABC}$ may be detected by Svetlichny inequality, which is given by \cite{svetlichny}
	\begin{eqnarray}
		|\langle S_{v}\rangle_{\rho_{ABC}}|\leq 4
		\label{svetineq}
	\end{eqnarray}
	where $S_{v}$ denotes the Svetlichny operator, which may be defined as
	\begin{eqnarray}
		S_{v}&=&\vec{a}.\vec{\sigma_{1}}\otimes [\vec{b}.\vec{\sigma_{2}}\otimes (\vec{c}+\vec{c'}).\vec{\sigma_{3}}+\vec{b'}.\vec{\sigma_{2}}\otimes (\vec{c}-\vec{c'}).\vec{\sigma_{3}}]\nonumber\\&+& \vec{a'}.\vec{\sigma_{1}}\otimes [\vec{b}.\vec{\sigma_{2}}\otimes (\vec{c}-\vec{c'}).\vec{\sigma_{3}}-\vec{b'}.\vec{\sigma_{2}}\otimes (\vec{c}+\vec{c'}).\vec{\sigma_{3}}]\nonumber \\
		%\label{svetop}
	\end{eqnarray}
	Here $\vec{a},\vec{a'}$, $\vec{b},\vec{b'}$, and
	$\vec{c},\vec{c'}$ are the unit vectors and the $\vec{\sigma_{i}}=(\sigma_{i}^{x},\sigma_{i}^{y},\sigma_{i}^{z})$ denote the spin projection operators.\\
	The expectation value of the Svetlichny operator with respect to the three-qubit state $\rho_{ABC}$ is given by \cite{mli,lysun}
	\begin{eqnarray}
		\langle S_{v} \rangle_{\rho_{ABC}}&=& Max_{\vec{a},\vec{b},\vec{c},\vec{a'},\vec{b'},\vec{c'}}([\vec{a}.\vec{\sigma_{1}}\otimes \vec{b}.\vec{\sigma_{2}}-\vec{a'}.\vec{\sigma_{1}}\otimes\vec{b'}.\vec{\sigma_{2}}]^{T}\nonumber\\&& M(\vec{c}+\vec{c'}).\vec{\sigma_{3}}
		+[\vec{a}.\vec{\sigma_{1}}\otimes \vec{b'}.\vec{\sigma_{2}}+\vec{a'}.\vec{\sigma_{1}}\otimes\vec{b}.\vec{\sigma_{2}}]^{T}\nonumber\\&&M(\vec{c}-\vec{c'}).\vec{\sigma_{3}})
		\label{svetop}
	\end{eqnarray}
	where $M=(M_{j,ik})$ represents a matrix with the entries $M_{ijk}={Tr\rm}(\sigma_{i}\otimes \sigma_{i} \otimes \sigma_{k})$, $i,j,k=1,2,3$.\\
	If any three-qubit state $\rho_{ABC}$ violates the inequality (\ref{svetineq}) then $\rho_{ABC}$ can be considered as a genuine tripartite non-local state.\\
	M. Li et al. \cite{mli} found that the expectation value of the Svetlichny operator $S_{v}$ with respect to any three-qubit state is bounded above, and it is given by
	\begin{eqnarray}
		|\langle S_{v}\rangle_{\rho_{ABC}}|\leq 4\mu_{1}
		\label{svetup}
	\end{eqnarray}
	where $\mu_{1}$ denotes the maximum singular value of the matrix $M$.\\
	%\subsection{Relationship between the non-locality of two-qubit state $\rho$ measured by $S_{NL}(\rho)$ and the non-locality of pure three-qubit state measured by Svetlichney operator}
	We are now in a position to establish a relationship between $S_{NL}(\rho_{AB})$ and $\langle S_{v}\rangle_{\rho_{ABC}}$. To do this, let us first consider a canonical form of a pure three-qubit state, which is given by \cite{acin}
	\begin{eqnarray}
		|\psi\rangle_{ABC}&=&\lambda_{0}|000\rangle_{ABC}+\lambda_{1}e^{i\theta}|100\rangle_{ABC}+\lambda_{2}|101\rangle_{ABC} \nonumber\\&+&\lambda_{3}|110\rangle_{ABC}+\lambda_{4}|111\rangle_{ABC},~\sum_{i=0}^{4}\lambda_{i}^{2}=1
		\label{can3state}
	\end{eqnarray} 
	where $0\leq \lambda_{i} \leq 1$ and $0\leq \theta \leq \pi$.\\
	To achieve the required relation, we take into account the two-qubit state described by the density operator $\rho_{AB}$, whose purification is the three-qubit state $|\psi\rangle_{ABC}$ \cite{laxmi}. The state $\rho_{AB}$ is given by
	\begin{eqnarray}
		\rho_{AB}=\begin{pmatrix} 
			\lambda_{0}^{2} & 0 & \lambda_{0}\lambda_{1}e^{i\theta} & \lambda_{0}\lambda_{3} \\
			0 & 0 & 0 & 0\\
			\lambda_{0}\lambda_{1}e^{-i\theta} & 0 & \lambda_{1}^{2}+\lambda_{2}^{2} & \lambda_{1}\lambda_{3}e^{-i\theta}+\lambda_{2}\lambda_{4}\\
			\lambda_{0}\lambda_{3} & 0 & \lambda_{1}\lambda_{3}e^{i\theta}+\lambda_{2}\lambda_{4} & \lambda_{3}^{2}+\lambda_{4}^{2} \\
		\end{pmatrix}\nonumber \\
		\label{two-qubitpuri}
	\end{eqnarray}
	We can make an observation that it is not very easy to obtain the analytical relationship between the expectation value of the Svetlichny operator with respect to the pure-three qubit state $|\psi\rangle_{ABC}$, and the strength of the non-locality of two-qubit mixed state $\rho_{AB}$ by keeping all the parameters. Thus to obtain the required relationship, we consider a few particular types of three-qubit states.
	\subsubsection{A family of pure three-qubit states: GHZ class}
	\noindent Let us consider a pure three-qubit state, which can be expressed as
	\begin{eqnarray}
		|\psi^{MS}\rangle_{ABC} &=&\frac{1}{\sqrt{2}} 
		(|000\rangle_{ABC}+cos\theta|110\rangle_{ABC}\nonumber\\&+& 
		sin\theta|111\rangle_{ABC}),~0< \theta < \frac{\pi}{2}
	\end{eqnarray}  
	It is known as the maximal slice (MS) state \cite{rungta}. 
	The inherent symmetries of the MS state make it very useful for quantum communication purposes \cite{sohini}. The expectation value of Svetlichny operator $S_{v}$ with respect to the state $|\psi^{MS}\rangle_{ABC}$ is given by \cite{rungta}
	\begin{eqnarray}
		\langle S_{v} \rangle_{|\psi^{MS}\rangle_{ABC}}&=&4\sqrt{2-Cos^{2}\theta}
		\label{sv123}
	\end{eqnarray}
	Using (\ref{two-qubitpuri}), we can obtain the two-qubit state described by the density operator $\rho^{MS}_{AB}$, whose purification is the state $|\psi^{MS}\rangle_{ABC}$. The state $\rho^{MS}_{AB}$ is given by
	\begin{eqnarray}
		\rho^{MS}_{AB}=\begin{pmatrix} 
			\frac{1}{2} & 0& 0&\frac{cos\theta}{2} \\
			0 & 0 & 0 & 0\\
			0 & 0 & 0 & 0\\
			\frac{cos\theta}{2} & 0& 0& \frac{1}{2}\\
		\end{pmatrix}
	\end{eqnarray}
	The negativity of the state $\rho^{MS}_{AB}$ is given by
	\begin{eqnarray}
		N(\rho^{MS}_{AB})=\sqrt{\frac{1+cos2\theta}{2}}
		\label{neg1}
	\end{eqnarray}
	Consider the Bell-CHSH witness operator $W_{CHSH}^{ij}~(i,j=x,y,z;i\neq j)$ in $xy-$, $yz-$ and $zx-$ plane to detect the state $\rho^{MS}_{AB}$. The witness operator $W_{CHSH}^{ij}$ is given by
	\begin{eqnarray}
		W_{CHSH}^{ij}=2I-B_{ij},~ i,j=x,y,z;i\neq j
		\label{wchsh12}
	\end{eqnarray}
	where $B_{ij}=\sqrt{2}[\sigma_{i}\otimes \sigma_{i}+\sigma_{j}\otimes \sigma_{j}],~i,j=x,y,z~   \text{and}~ i \neq j$.\\
	Let us now discuss different cases by considering the witness operator in different two-dimensional planes.\\
	\textbf{Case-I:} \textbf{$xy-$ plane}.\\
	The witness operator defined in this plane is given by
	\begin{eqnarray}
		W_{CHSH}^{xy}=2I-\sqrt{2}[\sigma_{x}\otimes \sigma_{x}+\sigma_{y}\otimes \sigma_{y}]
		\label{wchsh121}
	\end{eqnarray}
	The expectation value of $W_{CHSH}^{xy}$ with respect to the state $\rho^{MS}_{AB}$ is given by
	\begin{eqnarray}
		{Tr\rm}[W_{CHSH}^{xy}\rho^{MS}_{AB}]= 2>0, \forall~~\theta \in (0,\frac{\pi}{2})
		\label{wchsh122}
	\end{eqnarray}
	The witness operator $W_{CHSH}^{xy}$ does not detect the state $\rho^{MS}_{AB}$ for any value of $\theta \in (0,\frac{\pi}{2})$.\\
	Therefore, the strength of the non-locality of $\rho^{MS}_{AB}$ can be obtained as
	\begin{eqnarray}
		S_{NL}(\rho^{MS}_{AB})=q.( P_{xy}^{max}-\frac{3}{4})+(1-q).K
	\end{eqnarray} 
	where $K=\frac{{Tr\rm}[W_{CHSH}^{xy}.\rho^{MS}_{AB}.(\rho^{MS}_{AB})^{T_{B}}]}{4.N(\rho^{MS}_{AB})}=\frac{1}{4cos\theta}$ and $q<(0.5,1]$. Further, we have $ P_{max}^{xy}-\frac{3}{4}=-\frac{1}{4}$. Using these values, we can get the expression for the strength of the non-locality of $\rho^{MS}_{AB}$ as 
	\begin{eqnarray}
		S_{NL}(\rho^{MS}_{AB})= \frac{1-q(1+cos\theta)}{4cos\theta},~q<(0.5,1]
		\label{snl99}
	\end{eqnarray} 
	In particular, considering $q=0.3$, the expression for $S_{NL}(\rho^{MS}_{AB})$ given in (\ref{snl99}) reduces to
	\begin{eqnarray}
		S_{NL}(\rho^{MS}_{AB})= \frac{0.7-0.3cos\theta}{4cos\theta},~0<\theta< \frac{\pi}{2}
		\label{snl991}
	\end{eqnarray} 
	As $\theta$ varies from $0$ to $\frac{\pi}{2}$, $S_{NL}(\rho^{MS}_{AB})\in (0.1,1.2]$.\\
	Using (\ref{sv123}) and (\ref{snl991}), we obtain a relation between $S_{NL}(\rho^{MS}_{AB})$ and $\langle S_{v}\rangle_{|\psi^{MS}\rangle_{ABC}}$ as
	\begin{eqnarray}
		\langle S_{v}\rangle_{|\psi^{MS}\rangle_{ABC}}&=&4\sqrt{2-\frac{49}{1600}.\frac{1}{(S_{NL}((\rho^{MS}_{AB})+\frac{3}{40})^{2}}},\nonumber \\&&
		0.1 < S_{NL}(\rho^{MS}_{AB}) \leq 1.2 
	\end{eqnarray}
	and the values of $\langle S_{v}\rangle_{|\psi^{MS}\rangle_{ABC}}$ with respect to $S_{NL}(\rho^{MS}_{AB})$ in $xy$-plane are shown in figure-2.
	\begin{figure}[h!]
		\centering
		\includegraphics[scale=0.68]{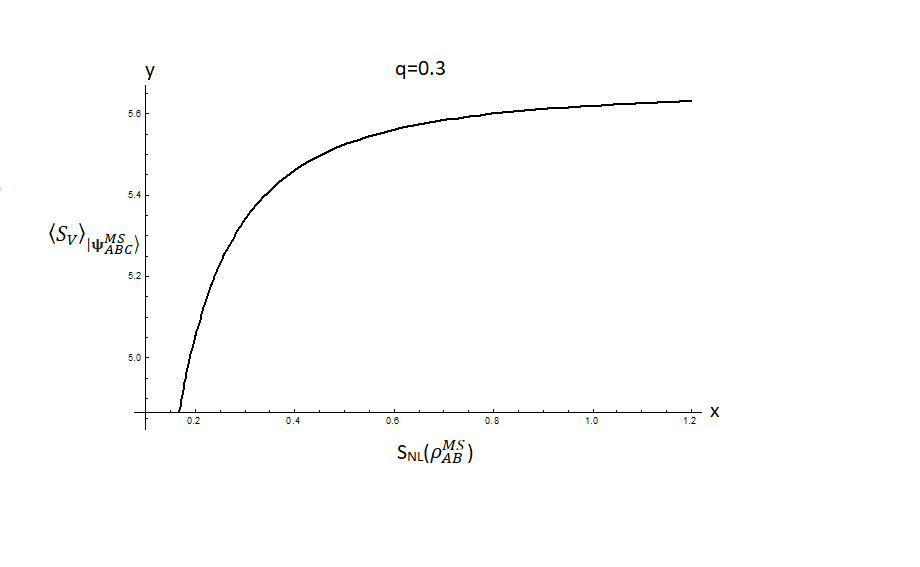}
		\caption{The graph depicts the relationship between  $\langle S_{v}\rangle_{|\psi^{MS}\rangle_{ABC}}$ and $S_{NL}(\rho^{MS}_{AB})$. It is clear from the graph that for $S_{NL}(\rho^{MS}_{AB})$ belongs to (0.1,1.2], $\langle S_{v}\rangle_{|\psi^{MS}\rangle_{ABC}}$ is always greater than 4, i.e., $S_{v}$ inequality is violated.}
	\end{figure}
	
	\textbf{Case-II:} \textbf{$yz-$ plane}.\\
	The witness operator defined in this plane is given by
	\begin{eqnarray}
		W_{CHSH}^{yz}=2I-\sqrt{2}[\sigma_{y}\otimes \sigma_{y}+\sigma_{z}\otimes \sigma_{z}]
		\label{wchsh131}
	\end{eqnarray}
	The expectation value of $W_{CHSH}^{yz}$ with respect to the state $\rho^{MS}_{AB}$ is given by
	\begin{eqnarray}
		\hspace*{-0.5cm}	{Tr\rm}[W_{CHSH}^{yz}\rho^{MS}_{AB}]= 2-\sqrt{2}+\sqrt{2}cos\theta>0,\forall~ \theta \in (0,\frac{\pi}{2})\nonumber \\
		\label{wchsh132}
	\end{eqnarray}
	In this case also, the witness operator $W_{CHSH}^{yz}$ does not detect the state $\rho^{MS}_{AB}$ for any value of $\theta \in (0,\frac{\pi}{2})$.\\
	Therefore, the strength of the non-locality of $\rho^{MS}_{AB}$ can be obtained as
	\begin{eqnarray}
		S_{NL}(\rho^{MS}_{AB})=q.( P_{yz}^{max}-\frac{3}{4})+(1-q).K
	\end{eqnarray} 
	where $K=\frac{{Tr\rm}[W_{CHSH}^{yz}.\rho^{MS}_{AB}.(\rho^{MS}_{AB})^{T_{B}}]}{4.N(\rho^{MS}_{AB})}=\frac{2-\sqrt{2}+\sqrt{2}cos\theta}{8cos\theta}$ and $q<(0.5,1)~ \text{for}$ $\theta \in (0,\frac{\pi}{2})$. Further, we have $ P_{max}^{yz}-\frac{3}{4}=-\frac{2-\sqrt{2}+\sqrt{2}cos\theta}{8}$. Using these values, we can get the expression for the strength of the non-locality of $\rho^{MS}_{AB}$ as 
	\begin{eqnarray}
		S_{NL}(\rho^{MS}_{AB})= f(\theta)
		\label{snl999}
	\end{eqnarray} 
	where $f(\theta)=\frac{2-2\sqrt{2}sin^{2}\frac{\theta}{2}-q(4cos^{2}\frac{\theta}{2}-\sqrt{2}Sin^{2}{\theta}))}{8cos\theta}$.\\
	In particular, considering $q=0.001$, the expression for $S_{NL}(\rho^{MS}_{AB})$ given in (\ref{snl999}) reduces to
	\begin{eqnarray}
		S_{NL}(\rho^{MS}_{AB})&=& -0.001(\frac{2-\sqrt{2}+\sqrt{2}cos\theta}{8})\nonumber\\&+& 0.999(\frac{2-\sqrt{2}+\sqrt{2}cos\theta}{8cos\theta}) 
		\label{snon-local}
	\end{eqnarray} 
	As $\theta \in (0,\frac{\pi}{2}]$, $S_{NL}(\rho^{MS}_{AB})\in (0.25,0.7]$.\\
	Using (\ref{sv123}) and (\ref{snon-local}), we obtain a relation between $S_{NL}(\rho^{MS}_{AB})$ and $\langle S_{v}\rangle_{|\psi^{MS}\rangle_{ABC}}$ as
	\begin{eqnarray}
		\langle S_{v}\rangle_{|\psi^{MS}\rangle_{ABC}}=\sqrt{32-u^{2}},
		~0.25 < S_{NL}(\rho^{MS}_{AB}) \leq 0.7 \nonumber
	\end{eqnarray}
	\\
	where $u=\frac{-(8S_{NL}-1.41221)+\sqrt{(8S_{NL}-1.41221)^2+0.00330521}}{0.007}$. The values of $\langle S_{v}\rangle_{|\psi^{MS}\rangle_{ABC}}$ with respect to $S_{NL}(\rho^{MS}_{AB})$ in $xy$-plane are shown in figure-3.\\
	In a similar fashion, we can obtain the relationship between $\langle S_{v}\rangle_{|\psi^{MS}\rangle_{ABC}}$ and $S_{NL}(\rho^{MS}_{AB})$ when the witness operator $W_{CHSH}$ defined in $zx-$ plane.
	\begin{figure}[h!]
		\centering
		\includegraphics[scale=0.65]{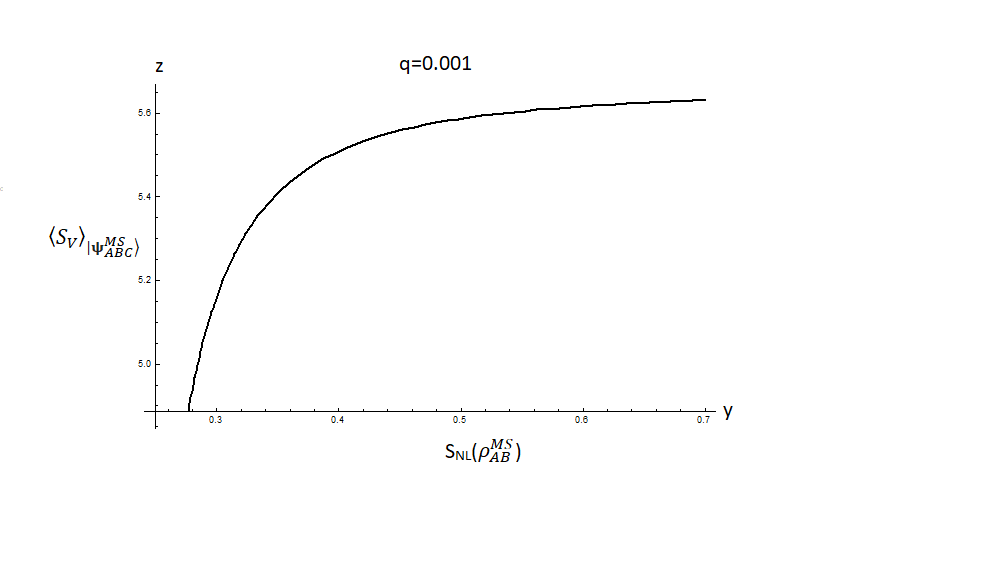}
		\caption{The graph depicts the relationship between  $\langle S_{v}\rangle_{|\psi^{MS}\rangle_{ABC}}$ and $S_{NL}(\rho^{MS}_{AB})$. It is clear from the graph that for $S_{NL}(\rho^{MS}_{AB})$ belongs to (0.25,0.7], $\langle S_{v}\rangle_{|\psi^{MS}\rangle_{ABC}}$ is always greater than 4, i.e., $S_{v}$ inequality is violated. }
	\end{figure}
	\subsubsection{A family of pure three-qubit states: W-class of Type-I}
	Let us consider a family of pure three-qubit states, which can be expressed in the form as 
	\begin{eqnarray}
		|\psi_{1}\rangle_{ABC}&=& \lambda_{0}|000\rangle_{ABC} +0.3|101\rangle_{ABC} \nonumber \\
		&&~~~+\sqrt{0.91-\lambda_{0}^{2}} |110\rangle_{ABC} \nonumber
	\end{eqnarray} 
	$|\psi_{1}\rangle_{ABC}$ represent a legitimate quantum state when $\lambda_{0}\in [0,0.953939]$. The state $|\psi_{1}\rangle_{ABC}$ belongs to the $W-$ class of states.\\
	Let us consider a two-qubit state described by the density operator $\rho^{(t1)}_{AB}$ which when purified, gives rise to the three-qubit pure state $|\psi_{1}\rangle_{ABC}$. The two-qubit state $\rho^{(t1)}_{AB}$ is given by \cite{laxmi}  
	\begin{eqnarray}
		\hspace*{-0.5cm}	\rho^{(t1)}_{AB}=\begin{pmatrix} 
			\lambda_{0}^{2} & 0& 0&\lambda_{0}\sqrt{0.91-\lambda_{0}^{2}} \\
			0 & 0 & 0 & 0\\
			0 & 0 & 0.09 & 0\\
			\lambda_{0}\sqrt{0.91-\lambda_{0}^{2}} & 0& 0& 0.91-\lambda_{0}^{2}\\
		\end{pmatrix}
	\end{eqnarray}
	$\text{where}~~\lambda_{0}\in [0,0.953939]$. In this interval of $\lambda_{0}$, the state $\rho^{(t1)}_{AB}$ is an entangled state, but it is not detected by the CHSH witness operators $W^{xy}_{CHSH}$ and $W^{yz}_{CHSH}$. The entangled state $\rho^{(t1)}_{AB}$ is only detected by the CHSH witness operator $W^{xz}_{CHSH}$. \\ 
	In the $xz-$ plane, the expectation value of CHSH witness operator $W^{xz}_{CHSH}$ with respect to the state $\rho^{(t1)}_{AB}$ is given by  
	\begin{eqnarray} 
		{Tr\rm}[W^{xz}_{CHSH}\rho^{(t1)}_{AB}]&=&0.840345 - 2.82843\lambda_{0}\sqrt{0.91-\lambda_{0}^{2}}\nonumber\\&<& 0,~~\lambda_{0}\in [0.335,0.85]
	\end{eqnarray}
	where $W^{xz}_{CHSH}=2I-B_{xz}~~ \text{and}$ 
	$B_{xz}=\sqrt{2}[\sigma_{x}\otimes \sigma_{x}+\sigma_{z}\otimes \sigma_{z}]$.\\ 
	Therefore, in this case, the non-locality of the two-qubit state 
	$\rho^{(t1)}_{AB}$ can be calculated via the formula
	\begin{eqnarray}
		S_{NL}(\rho_{AB}^{(t1)}) &=&-\frac{{Tr\rm}[W_{CHSH}^{xz}\rho^{(t1)}_{AB}]}{8}\nonumber\\ 
		&=& \frac{-(0.840345 - 2.82843\lambda_{0}\sqrt{0.91-\lambda_{0}^{2}})}{8}\hspace*{0.8cm}
		\label{snl400}
	\end{eqnarray}
	It can be easily found that the value of $S_{NL}(\rho_{AB}^{(t1)})$ lies in the interval $[0,0.06]$ when $\lambda_{0} \in [0.335,0.85].$ \\
	The expression (\ref{snl400}) can be re-expressed as 
	\begin{eqnarray}
		\lambda_{0}^{2}=\frac{0.91\pm \sqrt{(0.91)^{2}-k}}{2}
		\label{l0}
	\end{eqnarray}
	where $k= [\frac{8S_{NL}(\rho_{AB}^{(1)})+0.840305}{2.82843}]^{2}$.\\
	Now, our task is to calculate the expectation value of the Svetlichny operator with respect to the state $|\psi_{1}\rangle_{ABC}$. To accomplish this task, firstly, we need to calculate the matrix $M_{1}$ \cite{mli}, which is given by 
	\begin{eqnarray}
		M_{1}=\begin{pmatrix} 
			0 & 0& a& 0& 0&0& 0& 0&c  \\
			0 & 0 & 0 & 0&0&-a&0&0&0\\
			0 & 0 &b&0& 0 &0&0&0 &0.82\\
		\end{pmatrix}
	\end{eqnarray}
	where $a= 2\lambda_{0}\sqrt{0.91-\lambda_{0}^{2}},b=-0.6\sqrt{0.91-\lambda_{0}^{2}}~~ and ~~c= 0.6\lambda_{0}$. \\
	The maximum singular value of $M_{1}$ is given by
	\begin{eqnarray}
		\mu_{1}= 0.707107\sqrt{1+3.64\lambda_{0}^{2}-4\lambda_{0}^{4}+\sqrt{J}}
		\label{singmax}
	\end{eqnarray}
	where 
	$J=1-7.28\lambda_{0}^{2}+21.2496\lambda_{0}^{4}-29.12\lambda_{0}^{6}+16\lambda_{0}^{8}$\\
	Using the result (\ref{svetup}) and (\ref{singmax}), we get
	\begin{eqnarray}
		\langle S_{v}\rangle_{\rho_{ABC}^{1}}\leq 4(0.707107\sqrt{1+3.64\lambda_{0}^{2}-4\lambda_{0}^{4}+\sqrt{J}})\hspace*{1.0cm}
	\end{eqnarray}
	where $\rho_{ABC}^{1}=|\psi_{1}\rangle_{ABC}\langle \psi_{1}|$.\\
	When the state parameter $\lambda_{0}$ is given by (\ref{l0}), then the relation between $|\langle S_{v}\rangle_{\rho_{ABC}^{1}}|$ and $S_{NL}(\rho_{AB}^{(t1)})$ may be written as
	\begin{eqnarray}
		|\langle S_{v}(\rho_{ABC}^{1})\rangle|\leq 4(0.707107\sqrt{1+3.64\lambda_{0}^{2}-4\lambda_{0}^{4}+\sqrt{l}})\hspace*{1.0cm}
	\end{eqnarray}
	One can now easily verify that the pure three-qubit state $|\psi_{1}\rangle_{ABC}$ satisfies the Svetlichny inequality when $S_{NL}(\rho_{AB}^{(t1)}) \in [0,0.06]$.\\
	\subsubsection{A family of pure three-qubit states: W-class of Type-II}
	Consider a family of pure three-qubit state
	\begin{eqnarray}
		|\psi_{2}\rangle_{ABC}&=& \lambda_{0}|000\rangle_{ABC} +0.7|100\rangle_{ABC} \nonumber\\
		&&~~~ +\sqrt{0.51-\lambda_{0}^{2}} |110\rangle_{ABC} \nonumber
	\end{eqnarray}
	The state $|\psi_{2}\rangle_{ABC}$ is defined for $\lambda_{0}\in [0.1,0.7]$. The state $|\psi_{2}\rangle_{ABC}$ belongs to $W-$ class of states.\\
	The two-qubit state $\rho^{(t2)}_{AB}$ can be purified to $	|\psi_{2}\rangle_{ABC}$. The density matrix $\rho^{(t2)}_{AB}$ is given by \cite{laxmi}  
	\begin{eqnarray}
		\rho^{(t2)}_{AB}=\begin{pmatrix} 
			\lambda_{0}^{2} & 0& 0.7\lambda_{0} &\lambda_{0}t \\
			0 & 0 & 0 & 0\\
			0.7\lambda_{0} & 0 & 0.49 & 0.7 t \\
			\lambda_{0}t & 0& 0.7 t& t \\
		\end{pmatrix},~~\lambda_{0}\in [0.1,0.7]\hspace*{0.4cm}
	\end{eqnarray}
	where $t=\sqrt{0.51-\lambda_{0}^{2}}$. In the given interval of $\lambda_{0}$, the state $\rho^{(t2)}_{AB}$ is an entangled state, but it is not detected by any of the CHSH witness operators $W^{xy}_{CHSH}$, $W^{yz}_{CHSH}$, and $W^{xz}_{CHSH}$. Therefore, we can proceed with any one of the CHSH witness operators. Let us choose the witness operator $W^{xy}_{CHSH}$. In the $xy-$ plane, the expectation value of CHSH witness operator $W^{xy}_{CHSH}$ with respect to the state $\rho^{(t2)}_{AB}$ is given by  
	\begin{eqnarray} 
		\hspace*{-0.5cm}	{Tr\rm}[W^{xy}_{CHSH}\rho^{(t2)}_{AB}]= 2> 0,~~\lambda_{0}\in [0.1,0.7]
	\end{eqnarray}
	where $W^{xy}_{CHSH}=2I-B_{xy}~~ \text{and}$ 
	$B_{xy}=\sqrt{2}[\sigma_{x}\otimes \sigma_{x}+\sigma_{y}\otimes \sigma_{y}]$.\\ 
	Therefore, in this case, the non-locality of a two-qubit entangled state $\rho^{(t2)}_{AB}$ can be calculated as
	\begin{eqnarray}
		S_{NL}&=&q( P_{max}^{xy}-\frac{3}{4})+(1-q)k 
		\label{snl401} 
	\end{eqnarray} 
	where $k=2 -2.04\lambda_{0}^{2} + 4\lambda_{0}^{4} - 1.38593\lambda_{0}\sqrt{0.51 -\lambda_{0}^{2}}$ and $P_{max}^{xy}=\frac{1}{2}$.\\
	The parameter $q$ satisfies the inequality
	\begin{eqnarray}
		q<[0.73,1]\label{l000}
	\end{eqnarray}
	Considering $q=0.6$, the strength of the non-locality of $\rho_{AB}^{(t2)}$ is given by
	\begin{eqnarray}
		S_{NL}(\rho_{AB}^{(t2)})=\frac{1+2\lambda_{0}^{4}-K}{\lambda_{0}\sqrt{51-100\lambda_{0}^{2}}}
	\end{eqnarray} 
	where $K=[1.02\lambda_{0}^{2}+0.15\lambda_{0}\sqrt{51-100\lambda_{0}^{2}} +0.692965\lambda_{0}\sqrt{0.51-\lambda_{0}^{2}}]$.\\
	It can be easily seen that the value of $S_{NL}(\rho_{AB}^{(t2)})\in [0.1219,1.18077]$ for $\lambda_{0}\in[0.1,0.7]$. \\
	For the state $|\psi_{2}\rangle_{ABC}$, the matrix $M_{2}$ is given by \cite{mli} 
	\begin{eqnarray}
		M_{2}=\begin{pmatrix} 
			0 & 0& a_{1}& 0& 0&0& b_{1}& 0&0  \\
			0 & 0 & 0 & 0&0&-a_{1}&0&b_{1}&0\\
			c_{1} & 0 & 0 & 0 & -c_{1} &0&0&0 &1\\
		\end{pmatrix}
	\end{eqnarray}
	where $a_{1}= 2\lambda_{0}\sqrt{0.51-\lambda_{0}^{2}}, b_{1}=-1.4\sqrt{0.51-\lambda_{0}^{2}}$ and $c_{1}= 1.4\lambda_{0}$.\\
	The maximum singular value of $M_{2}$ is given by
	\begin{eqnarray}
		\mu_{2}= \sqrt{1+3.92\lambda_{0}^{2}}
		\label{singmax1}
	\end{eqnarray}
	Using the result (\ref{svetup}) and (\ref{singmax1}), we get
	\begin{eqnarray}
		\langle S_{v}\rangle_{\rho'_{ABC}}\leq 4 \sqrt{1+3.92\lambda_{0}^{2}} \hspace*{1.0cm}
	\end{eqnarray}
	where $\rho'_{ABC}=|\psi_{2}\rangle_{ABC}\langle \psi_{2}|$.\\
	The relation between $|\langle S_{v}\rangle_{\rho'_{ABC}}|$ and $S_{NL}(\rho_{AB}^{(2)})$ may be given by
	\begin{eqnarray}
		4 < \langle S_{v}(\rho'_{ABC})\rangle \leq  \sqrt{16+\frac{33.1546}{S_{NL}(\rho_{AB}^{t2})}}\nonumber
	\end{eqnarray}
	One can now find that the pure three-qubit state $|\psi_{2}\rangle_{ABC}$ violates the Svetlichny inequality, when $S_{NL}(\rho_{AB}^{(t2)}) \in [0.1219,1.18077]$.
	\subsection{Upper bound of the power of the controller in controlled quantum teleportation in terms of $S_{NL}$}
	Controlled quantum teleportation \cite{karlson} is a variant of quantum teleportation protocol \cite{bene}, where a party ccontrols the fidelity of the quantum teleportation. To explain the controlled quantum teleportation, let us consider a three-qubit state described by the density operator $\rho_{CAB}$, which is shared between three distant parties Alice, Bob, and Charlie. Alice and Bob possess the qubit $A$ and $B$, while the qubit $C$ is with Charlie. In the controlled quantum teleportation, Charlie performs measurement on his qubit C, and as a result, Alice and Bob share a two-qubit state described by the density operator $\rho_{AB}$. Alice and Bob then use the state $\rho_{AB}$ as a resource state to teleport a qubit. The state $\rho_{AB}$ contains Charlie's measurement parameter, and this parameter is also visible in the expression of the fidelity of teleportation. Thus, Charlie may control the teleportation fidelity by choosing the measurement parameter, and hence he may act as a controller in the teleportation protocol. To quantify Charlie's strength, one may define the power of the controller. To study the controller's power in controlled teleportation, we need to consider the two quantities: (i) Conditioned fidelity denoted by $F_{C}^{(C)}$, which is assumed to be greater than $\frac{2}{3}$, and (ii) Non-conditioned fidelity denoted by $F_{NC}$, which is assumed to be less than $\frac{2}{3}$. Therefore, the power denoted by $P^{(C)}$ may be defined as \cite{sohini,lee,artur}
	\begin{eqnarray}
		P^{(C)}= F_{C}^{(C)}-F_{NC}
		\label{p1}
	\end{eqnarray}	   
	In this section, we will show that the controller's power in the controlled quantum teleportation is upper bounded by the quantity $M(\rho_{AB})$ and hence the quantity $S_{NL}(\rho_{AB})$. To obtain the required results, we need to state two lemmas which are given below:\\
	\textbf{Lemma-3}: If $\tau$ denotes the tangle of the three-qubit pure state described by the density matrix $\rho_{CAB}$ and $N(\rho_{AB})$ denotes the negativity of the two-qubit state $\rho_{AB}={Tr\rm}_{C}(\rho_{CAB})$, then the conditioned fidelity $F_{C}^{(C)}$ is given by
	\begin{eqnarray}
		\frac{2}{3}< F_{C}^{(C)} \leq \frac{2}{3}+\frac{\sqrt{\tau+(\sqrt{2}\sqrt{N^{2}(\rho_{AB})+N(\rho_{AB})}-N(\rho_{AB}))^{2}}}{3}\nonumber \\
		\label{FCT}
	\end{eqnarray} 
	\textbf{Proof:} The conditioned fidelity $F_{C}^{(C)}$ is given by \cite{leejoo}
	\begin{eqnarray}
		F_{C}^{(C)}=\frac{2+\tau_{AB}}{3}=\frac{2+\sqrt{\tau+(C(\rho_{AB}))^{2}}}{3}
		\label{FCT1}
	\end{eqnarray} 
	where $\tau_{AB}$ denotes the partial tangle and it can be expressed in terms of $\tau$ as $\tau_{AB}=\sqrt{\tau+(C(\rho_{AB}))^{2}}$ \cite{leejoo}.\\
	Verstraete et al. \cite{Vers} proved that the lower bound of the negativity $(N(\rho_{AB}))$ of any two-qubit state $\rho_{AB}$ can be expressed as a function of the concurrence $C(\rho_{AB})$ of the state, and it is given by
	\begin{eqnarray}
		N(\rho_{AB})\geq \sqrt{(1-(C(\rho_{AB}))^{2}+C(\rho_{AB})^2}-(1-C(\rho_{AB}))\nonumber \\
		\label{FCT2}
	\end{eqnarray}
	Simplifying (\ref{FCT2}) and writing $C(\rho_{AB})$ in terms of $N(\rho_{AB})$, we get
	\begin{eqnarray}
		0\leq C(\rho_{AB})\leq -N(\rho_{AB})+\sqrt{2}\sqrt{N^{2}(\rho_{AB})+N(\rho_{AB})}\nonumber \\
		\label{FCT3}
	\end{eqnarray}
	Using (\ref{FCT3}) in (\ref{FCT1}) for the state $\rho_{AB}$, we get the upper bound of $F_{C}^{(C)}$. Furthermore, from (\ref{FCT1}), it is clear that $F_{C}^{(C)}>\frac{2}{3}$. Hence the lemma.\\
	\textbf{Lemma-4}: If $\rho_{CAB}$ denotes the three-qubit pure state, then the non-conditioned fidelity $F_{NC}$ is given by \cite{horod}
	\begin{eqnarray}
		F_{NC} \geq \frac{3+M(\rho_{AB})}{6}
		\label{FNC1}
	\end{eqnarray} 
	where $\rho_{AB}={Tr\rm}_{C}(\rho_{CAB})$ is the two-qubit mixed state shared between two distant parties as a resource state to execute the teleportation protocol.\\
	\textbf{Result-8}: If $\tau$ denotes the tangle of a three-qubit pure state described by the density matrix $\rho_{CAB}$ and if $P^{(C)}$ denotes the power of the controller in controlled teleportation, then the upper bound of the power is given by
	\begin{eqnarray}
		P^{(C)}&\leq& (\frac{1-M(\rho_{AB})}{6})\nonumber \\ &+&\frac{\sqrt{\tau+(\sqrt{2}\sqrt{N^{2}(\rho_{AB})+N(\rho_{AB})}-N(\rho_{AB}))^{2}}}{3} \nonumber \\
		\label{CP}
	\end{eqnarray}	
	\textbf{Proof:} The power $P^{(C)}$ of the controller can be re-written as
	\begin{eqnarray}
		P^{(C)}= F_{C}^{(C)}-F_{NC}
		\label{p1}
	\end{eqnarray}	
	Using Lemma-3 and Lemma-4, the power $P^{(C)}$ given in (\ref{p1}) reduces to the following inequality
	\begin{eqnarray}
		P^{(C)}&\leq& \big(\frac{2}{3}+\frac{\sqrt{\tau+(\sqrt{2}\sqrt{N^{2}(\rho_{AB})+N(\rho_{AB})}-N(\rho_{AB}))^{2}}}{3}\nonumber \\&-&(\frac{3+M(\rho_{AB})}{6})\big)\nonumber \\
		&=& \big((\frac{1-M(\rho_{AB})}{6})\nonumber \\&+&\frac{\sqrt{\tau+(\sqrt{2}\sqrt{N^{2}(\rho_{AB})+N(\rho_{AB})}-N(\rho_{AB}))^{2}}}{3}\big)
		\label{p2}
	\end{eqnarray}	
	Hence proved.\\
	Since it is assumed that $F_{C}^{(C)}>\frac{2}{3}$ and $F_{NC}<\frac{2}{3}$, so the power $P^{(C)}$ of the controller cannot be negative \cite{sohini,artur}. Thus, we may note the following:\\
	\textbf{Note-1:} If the two-qubit reduced state $\rho_{AB}$ does not violate the CHSH inequality, then $M(\rho_{AB})\leq 1$, and thus the non-conditioned fidelity $F_{NC}$ will be less than $\frac{2}{3}$. Hence, the power $P^{(C)}$ is always positive.\\
	\textbf{Note-2:} If the two-qubit reduced state $\rho_{AB}$ does violate the CHSH inequality, then $M(\rho_{AB})> 1$ and, in this case, the non-conditioned fidelity $F_{NC}> \frac{2}{3}$. Thus there may be a chance to get the negative power, which is not acceptable. But if we impose restriction on $M(\rho_{AB})$, then we can make the power positive. Hence, the power $P^{(C)}$ is positive, only when the following conditions hold
	\begin{eqnarray}
		1< M(\rho_{AB})< 1+2\sqrt{L}
	\end{eqnarray}
	where L=$\tau+(\sqrt{2}\sqrt{N^{2}(\rho_{AB})+N(\rho_{AB})}-N(\rho_{AB}))^{2}$.\\
	\textbf{Result-9}: If the reduced entangled state $\rho_{AB}$ violates the CHSH inequality and is detected by the witness operator $W_{CHSH}$ then the connection between the non-locality of $\rho_{AB}$ determined by $S_{NL}(\rho_{AB})$ and the three-qubit tangle $\tau$ is given by
	\begin{eqnarray}
		S_{NL}(\rho_{AB})< \frac{\sqrt{1+2\sqrt{L}}-1}{4}
		\label{snl456}
	\end{eqnarray}
	Now we are in a position to express the controller's power in terms of $S_{NL}(\rho_{AB})$. \\
	\textbf{Result-10}: Let us consider a three-qubit state $\rho_{CAB}$ shared between three parties, Alice, Bob, and Charlie. If the reduced entangled state $\rho_{AB}={Tr\rm}_{C}(\rho_{CAB})$ violates the CHSH inequality and is detected by the witness operator $W_{CHSH}$, then the  controller's power $P^{(C)}$ can be determined by  $S_{NL}(\rho_{AB})$, which is given by
	\begin{eqnarray}
		P^{(C)}<\frac{1}{6}-\frac{4}{3}(S_{NL}(\rho_{AB})(1+2S_{NL}(\rho_{AB}))
		\label{p123}
	\end{eqnarray} 
	\textbf{Proof:} Since $S_{NL}(\rho_{AB})\geq 0$ so, the upper limit of $S_{NL}(\rho_{AB})$ given in (\ref{snl456}) must be positive. This gives
	\begin{eqnarray}
		\frac{\sqrt{1+2\sqrt{L}}-1}{4}\geq 0
		\implies L < \frac{1}{4}
		\label{cond1}
	\end{eqnarray}
	Also, the inequality that establishes the relation between the controller's power $P^{(C)}$ and $S_{NL}(\rho_{AB})$ is given by
	\begin{eqnarray}
		P^{(C)}<\frac{\sqrt{L}}{3}-\frac{4S_{NL}(\rho_{AB})(1+2S_{NL}(\rho_{AB}))}{3}
		\label{cond2}
	\end{eqnarray} 
	Using the inequality (\ref{cond1}) in (\ref{cond2}), we get the required result.
	
	\section{Conclusion}
	To summarize, we have considered the problem of detection of non-locality of a given two-qubit state. It is now an accepted fact that non-locality and entanglement are two different concepts, and thus if a two-qubit state is entangled, then it is not necessary that it also depicts the non-local feature. Therefore, one can find many entangled states in the literature that may satisfy Bell's inequality. In the context of the detection of non-local property of a two-qubit entangled state, we consider a Bell game where the maximum probability $P^{max}$ of winning of the game is related to the expectation value of the Bell operator. We have defined the strength of non-locality $S_{NL}$ in terms of $P^{max}$ and, later on, re-expressed the expression of $S_{NL}$ in terms of witness operator. First, we made a connection between the strength of the non-locality and the CHSH witness operator and then discussed the estimation of the non-locality of the given entangled state in both cases when (i) the CHSH witness operator detects the entangled state and (ii) CHSH witness operator does not detect the entangled state. Also, we construct an inequality that gives the upper bound of the strength of the non-locality, and the upper bound is given in terms of the optimal witness operator. By doing this, we are able to detect the non-locality in the given two-qubit entangled state, which are undetected earlier by the Bell-CHSH operator. Furthermore, we also developed an interconnection between the strength of the non-locality of the two-qubit state and the expectation value of the Svetlichney operator with respect to a pure three-qubit state. This link paves the way to study the non-locality of a pure three-qubit state in terms of the non-locality of a two-qubit system. We are now on the way to develop the relationship between the non-locality of the two-qubit system and the three-qubit (pure and mixed) system. As an application, we have shown that the power of the controller in the controlled quantum teleportation is limited by the non-locality of the two-qubit state $\rho_{AB}$ determined by $S_{NL}(\rho_{AB})$, where $\rho_{AB}$ denote the reduced density operator of the pure three-qubit state $\rho_{CAB}$.

	%Considering again, Equation-(9), can also be rewritten as 
	%\begin{eqnarray}
	%	S_{NL}(\rho_{BC}^{2}) \leq \frac{1+2\lambda_{0}^{4}}{\lambda_{0}\sqrt{51-100\lambda_{0}^{2}}}\\
	%	~\nonumber\\
	%	\mbox{Since,~~~} 0<\lambda_{0}^{2}<0.51,\nonumber\\
	%	\mbox{thus,~~~} 1+2\lambda_{0}^{4}<1.4802, \nonumber\\  \mbox{and,~~~}\frac{1}{\sqrt{51-100\lambda_{0}^{2}}}\leq 0.7071.\nonumber
	%\end{eqnarray}
	
	%After rewriting Equation-(12), we get \begin{eqnarray}
		%	\frac{1}{\lambda_{0}}\leq \frac{16 \sqrt{3.92}}{\langle %S_{v}(\rho_{ABC}^{2})\rangle-16}
		%\end{eqnarray}
		%After putting the above expressions in Equation-(13), we obtain a relation between $\langle S_{v}(\rho_{ABC}^{2})\rangle$ and $S_{NL}(\rho_{BC}^{2})$ when $\langle S_{v}(\rho_{ABC}^{2})\rangle >4$ and which is given as
		%\begin{eqnarray}
		%	4 < \langle S_{v}(\rho_{ABC}^{2})\rangle \leq  %\sqrt{16+\frac{33.1546}{S_{NL}(\rho_{BC}^{2})}}\nonumber
		%\end{eqnarray}
		%And by using the above expression $S_{v}(\rho_{ABC}^{2})\rangle$ lies between [7,17] when $ S_{NL}(\rho_{BC}^{2}) $ lies between [0.1219,1.18077].
		
		\section{Acknowledgement}
		\noindent A. G. would like to acknowledge the financial support from CSIR. This work is supported by CSIR File No. 08/133(0035)/2019-EMR-1.
		
		\section{DATA AVAILABILITY STATEMENT}
		Data sharing not applicable to this article as no datasets were generated or analysed during the current study.

	\end{document}